\documentclass[fleqn,usenatbib]{mnras}

\usepackage{newtxtext,newtxmath}

\usepackage[T1]{fontenc}

\usepackage{graphicx}	
\usepackage{amsmath}	
\usepackage{comment}
\usepackage{tikz}
\usepackage{epsfig}           
\usepackage{lscape}    
\usepackage{booktabs} 

\title[Interference detection in radio astronomy]{Interference detection in radio astronomy applying Shapiro-Wilks normality test, spectral entropy, and spectral relative entropy}

\author[Z. Cao, et al.]{Zhicheng Cao$^{1}$\thanks{E-mail: zccao@xidian.edu.cn},
Natalia A. Schmid$^{2,3}$\thanks{E-mail: Natalia.Schmid@mail.wvu.edu (Corresponding author)},
    Kevin Bandura$^{2,3}$\thanks{E-mail: Kevin.Bandura@mail.wvu.edu},
    Duncan R. Lorimer$^{3,4}$\thanks{E-mail: Duncan.Lorimer@mail.wvu.edu},
    Morgan Dameron$^{2}$, 
\newauthor
    Katelyn Crockett$^{2}$,
    Clayton Grubick$^{2}$, 
    Andreas Schmid$^{3,5}$ and
    Shaonan Zheng$^{1}$
\\
$^{1}$ Xidian University, School of Life Science and Technology, Shaanxi, 710126, China.\\   
$^{2}$ Lane Department of Computer Science and Electrical Engineering, West Virginia University, Morgantown, WV, USA.\\
$^{3}$ Center for Gravitational Waves and Cosmology, West Virginia University, Chestnut Ridge Research Building, Morgantown, WV, USA\\
$^{4}$ Department of Physics and Astronomy, West Virginia University, Morgantown, WV, USA\\
$^{5}$ Department of Electrical and Computer Engineering, University of Illinois in Urbana-Champaign, Urbana, IL USA\\
}

\date{Accepted 05/08/2024. Received 16/06/2024; in original form 06/01/2024}

\pubyear{2023}

\begin{document}
\label{firstpage}
\pagerange{\pageref{firstpage}--\pageref{lastpage}}
\maketitle 

\begin{abstract}  
Radio-frequency interference (RFI) is becoming an increasingly significant problem for most radio telescopes. Working with Green Bank Telescope data from PSR J1730+0747 in the form of complex-valued channelized voltages and their respective high-resolution power spectral densities, we evaluate a variety of statistical measures to characterize RFI. As a baseline for performance comparison, we use median absolute deviation (MAD) in complex channelized voltage data and spectral kurtosis (SK) in power spectral density data to characterize and filter out RFI. From a new perspective, we implement the Shapiro-Wilks (SW) test for normality and two information theoretical measures, spectral entropy (SE) and spectral relative entropy (SRE), and apply them to mitigate RFI.  The baseline RFI mitigation algorithms are compared against our novel RFI detection algorithms to determine how effective and robust the performance is. Except for MAD, we find significant improvements in signal-to-noise ratio through the application of SE, symmetrical SRE, asymmetrical SRE, SK, and SW. These algorithms also do a good job of characterizing broadband RFI. Time- and frequency-variable RFI signals are best detected by SK and SW tests.
\end{abstract} 




\section{Introduction} 
\label{sec:introduction}

\begin{figure*}
\centering 
\includegraphics[width=\textwidth]{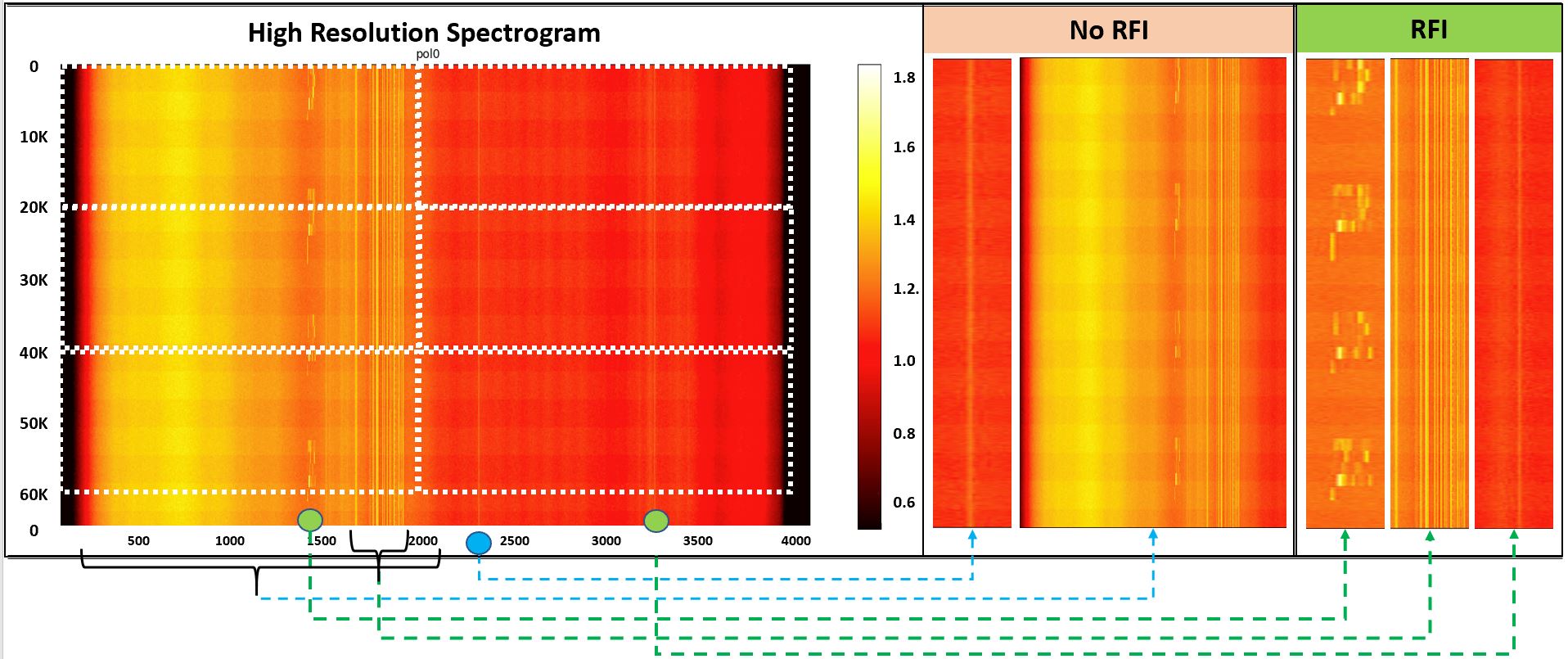}
\caption{ Snapshot of a sky-view high-resolution spectrogram (left), with the frequency channel along the $x$-axis ($0 = 1900$~MHz, $4096 = 1100$~MHz), and time in $y$-axis (0--65,024~time samples or 0--0.33~s), increasing in the downwards direction. On the right, we see, from left to right, two cases of no RFI (should not be flagged), namely ``Milky Way Galaxy'' around the discrete frequency channel 2300, a ``representation of a band-pass shape of a long bandwidth'' from the frequency channel 200 to the frequency channel 2100, and three cases of RFI (should be flagged), namely, the ``Iridium SatCom signals" in the frequency channels 1402-1433 demonstrating periodicity, the unknown RFI in the range of discrete frequency channels from 1700 to 1900, and the ``Bedford Radar'' in the frequency channel 3300. }
\label{figure:RFI_example}
\end{figure*}

Radio-frequency interference (RFI)  are electromagnetic signals negatively impacting  radio astronomical measurements. Both natural phenomena such as lightning strikes or the northern and southern lights and man-made devices such as radars, radio, television, cell phones, and satellites generate sources of RFI. Most of them are caused by using commodities as simple as a wireless telephone, an automotive radar installed on a car, or an aerial device flying close to the observatory. RFI may also be caused by failing electronics or by an open microwave located somewhere in a zone surrounding the telescope and leaking a radio signal. The amount of man-made RFI continues to increase as technology advances. For a recent review, see \citet{saroff}. As an illustration, Fig.~\ref{figure:RFI_example} shows several types of RFI typical for radio astronomy data. 
Currently, methods of RFI detection and removal are limited to the type of RFI, the position in which the excision algorithm is applied during the processing pipeline, and a radio telescope's hardware set-up \citep[see, e.g.,][]{Ford2014}. As the raw data are often  averaged  before any astronomical analysis, RFI becomes more capable of easily suppressing astronomical signals of interest and making them harder to study \citep{Real-timeRFI:eramey}.         

\begin{figure}
\centering
\includegraphics[width=0.5\textwidth]{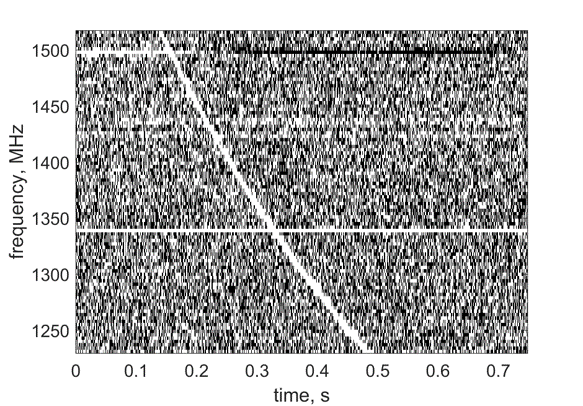}
\caption{Example data set showing the `Lorimer burst' (FRB~010724) in a frequency vs time plot. The white line sweeping left to right is the signal. The pixelated black and white background is noise being received at the same time as the pulse. Note the pulse is dispersed: the higher frequency components arrive earlier than their lower frequency counterparts.}
\label{fig:LorimerBurst}
\end{figure}

In this work, we examine the excision of RFI
from astronomical observations of transient phenomena in the radio sky. Examples of these sources are pulsars \citep{Lorimer12}, Rotating Radio Transients \citep[RRATs;][]{mclaughlin2006transient}, and Fast Radio Bursts \citep[FRBs;][]{lbm+07,2013Sci...341...53T}. 
Observations of these sources are most commonly done by collecting so-called filterbank data (power spectral density of astronomical data displayed as a function of observing time and sky frequency). An example pulse is shown for the first FRB in Figure~\ref{fig:LorimerBurst}.

Electromagnetic radiation from pulsars, RRATs, and FRBs arrive on Earth as extremely weak broadband signals. As an extraterrestrial signal propagates through space, it passes through an  environment, called the interstellar medium  (ISM), which is full of free electrons. This causes the signal to become dispersed. As shown in Figure~\ref{fig:LorimerBurst}, the result of dispersion is that the lower frequency components of the signal get delayed from the higher frequency components. The time delay observed,
\begin{equation}
\Delta t \simeq 4.15~{\rm ms} \times 
\left(\frac{\rm DM}{{\rm cm}^{-3}~{\rm pc}}\right) \times 
\left[
\left(\frac{f_{\rm low}}{\rm GHz}\right)^{-2} - 
\left(\frac{f_{\rm high}}{\rm GHz}\right)^{-2}
\right],
\label{eq:timeDelay}
\end{equation}
where the dispersion measure, DM, is the integrated column of free electrons over the line of sight and $f_{\rm low}$ and $f_{\rm high}$ are, respectively, the low and high frequencies of the received band. This unique dispersion property separates celestial signals from other signals. 

By the time the signal is received on Earth, its signal strength,
$x_{\rm source}$, has decreased dramatically \citep[typical power densities in the range --150 to --220~dBWm$^{-2}$, see e.g.,][]{Ford2014} and it must compete with noisy signals produced from the instrumentation and thermal background of the receiver, $x_{\rm system}$. In addition, any RFI that was transmitted across the same radio spectrum is also received as $x_{\rm RFI}$. The resulting amplitude of the signal is a sum 
\begin{equation}
x(t) = x_{\rm source}(t) + x_{\rm system}(t) + x_{\rm RFI}(t),
\label{eq:signalModel}
\end{equation}
where each component is a function of time, $t$. {Even in very remote sites, terrestrial and orbital RFI signals can dominate the astronomical signal. New and improved RFI detection and characterization approaches will fully utilize the sensitivity of radio telescopes. }

The goal of this research is to develop novel, high-level, and efficient, real-time RFI detection and flagging algorithms using inferential statistics and information theoretical measures in application to raw channelized voltages. In this paper, we will be working primarily with output from the Green Bank Telescope (GBT). On raw complex-valued channelized data, we explore the applications of symmetric and asymmetric spectral relative entropy 
\citep[SRE;][]{2011arXiv1103.5602F}, spectral entropy \citep[SE;][]{Shen1998RobustEE}, and Shapiro-Wilks (SW) test for normality \citep{Shapiro1965}. We will generate the resultant masks for each test. Since masks are generated through a thresholding procedure, different values of the threshold will be involved in testing to aid in determining the most effective value of the threshold. As a baseline for comparison with our algorithms, we will use two well-known RFI detection algorithms, spectral kurtosis \citep[SK;][]{Dwyer1983} and Median Absolute Deviation \citep[MAD;][]{Buch2016}. 

The main contributions from our work are fourfold: (i) we propose using SE, SRE, and the SW test for normality as new methods of RFI detection in raw complex-valued channelized voltage data; (ii) the main constraint of our design is that channels must be processed independently for the benefit of parallel implementation in FPGA or GPU; (iii) we compare the performance of the proposed methods to that of MAD and SK and illustrate the benefit of applying each method to the channelized voltage data of PSR~J1713+0747; (iv) we analyze the performance of each method by generating folded pulse profiles of PSR~J1713+0747 from the data and evaluating its signal-to-noise ratio (S/N).     

The remainder of this paper is organized as follows. Section \ref{sec:techniques} discusses the current state-of-the-art techniques for RFI detection and mitigation. In addition, it explains the basic foundations of inferential statistics and information theory techniques that were researched and developed for efficient, real-time RFI detection. Section \ref{sec:data} presents characteristics of the test data. Section \ref{sec:results} compares the observational and qualitative results of the various algorithms explored. It presents the results of different threshold values for RFI mask generation and analyzes the S/N of each method tested. Finally, in Section \ref{sec:summary} we summarize the main findings of our work and also provide suggestions for future research.

\section{RFI Detection and Mitigation Techniques} 
\label{sec:techniques}
There are many different RFI detection and mitigation methods currently implemented for radio telescopes. The detection or mitigation technique used varies greatly depending on the type of interference, hardware implementation, and the processing pipeline step in which the excision method is applied \citep{Ford2014}. However, not all excision methods are published. Moreover, they are often specific to the application the radio telescope is being used for, i.e. pulsar searches, FRBs searches, galaxy mapping, etc.

\subsection{Processing pipeline}
A radio telescope's receiver outputs data in time series, complex-valued voltages. These voltages are then converted to complex-valued channelized voltages where they are broken down into $K$ frequency channels and $N$ time samples (also called time bins). This is done by performing a short-time Fourier Transform (FT) over the time-series data. Here, each frequency channel represents a small portion of the receiver bandpass. Next, the pixel-wise power of the complex-valued channelized voltages is computed to create high-resolution power spectral density data known in radio astronomy as filter bank data \citep{Lorimer12}. In computer science, it is known as a spectrogram \citep{Flanagan1972}. Post-processing algorithms are applied to the complex-valued channelized voltages and spectrograms to sort the data and find astronomical signals of importance.

\subsection{Current state-of-the-art RFI mitigation techniques}
\label{sec:other_methods}
\begin{figure}
        \centering
        \includegraphics[width=0.45\textwidth]{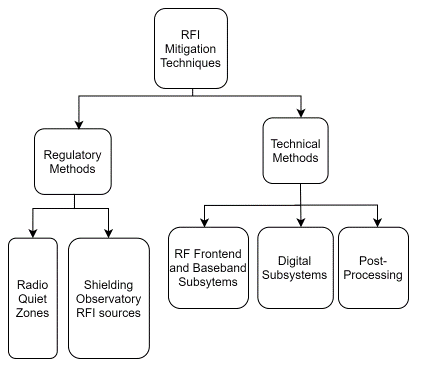}
        \caption{A breakdown of various places RFI mitigation can be performed as described by \citep{Ford2014}.}                                 
        \label{fig:mitigationBreakdown}
\end{figure}

RFI can be mitigated in a variety of locations in the observatory pipeline. This includes regulatory methods which are applied before a signal is received at a radio telescope and technical processing methods which are applied at various locations throughout the receiver's pipeline \citep{Ford2014}. A breakdown of each of these categories is shown in the block diagram in Figure~\ref{fig:mitigationBreakdown}.
\vskip \medskipamount
\noindent
{\bf Attenuation of terrestrial RFI}
Before technical mitigation methods are applied, observatories take regulatory methods to negate the effects of RFI. These efforts start with the locations radio observatories are built. They are strategically placed in sparse population density areas so that the narrowband RFI produced by man-made devices can be minimized. In the United States, for example, a National Radio Quiet Zone (NRQZ) exists for this purpose. First established on November 19, 1958, by the Federal Communications Commission and the Inter-department Radio Advisory Committee on March 26, 1958, the NRQZ was formed to minimize possible harmful interference with the NRAO and the United States Navy. The NRQZ covers roughly 13,000 square miles of land\footnote{https://greenbankobservatory.org/about/national-radio-quiet-zone} across West Virginia and Virginia, encompassing Green Bank Observatory (GBO) in Green Bank, West Virginia. For additional attenuation, electromagnetic shields, such as Faraday cages, are placed on-site around equipment and enclosures that emit electromagnetic leakage \citep{Ford2014}. However, the control over terrestrial RFI diminishes as more equipment is used at observatories. This increases the importance of RFI mitigation from other positions in the radio telescope pipeline. Analog RFI excision is performed in the receiving system of the telescope. It is at this point that signal processing and learning excision methods can be applied~\citep{Ford2014}.

\vskip \medskipamount
\noindent
{\bf Edge-thresholding}
This refers to a method to flag RFI against FRBs~\citep{Boyle2019}. It uses two unique characterization differences to flag regions of non-smooth, narrow, high-intensity data. First, it takes into account that FRBs are wider than RFI. Second, FRBs are pseudo-normally distributed. During edge-thresholding, data are processed iteratively across increasing window sizes. RFI becomes flagged when the difference between the window boundary and sample point is above a threshold, $T$, typically based on standard deviation or median absolute deviation. The algorithm is summarized by the decision rule
\begin{equation}
f(x_{i}) = {\rm min}(|x_{i} - x_{0}|, |x_{i}-x_{w}|) > T,
\label{eq:edgeThresh}
\end{equation}
where the data window $w = (x_{0},...,x_{w})$ and a point $x_{i} \in (x_{1}, x_{w-1})$ are flagged as RFI if they are greater than the set threshold $T$~\citep{Boyle2019}.

\vskip \medskipamount
\noindent
{\bf Spectral kurtosis (SK)}
This is a well-known method for the analysis of non-stationary non-Gaussian signals. Its initial development \citep{Dwyer1983} was applied to improve the detection of distorted underwater acoustic signals. It was later applied to radio astronomical data by \citet{nita2007} and is being increasingly 
used in this field to mitigate RFI. In terms of the principle of its operation, SK is based on the estimation of the fourth central moment known in probability theory as kurtosis in application to the data in the form of power spectral density (PSD). \citet{nita2007} showed that SK is a robust estimator to distinguish Gaussian noise from non-Gaussian RFI using power spectral density data. It is based on a selection of $M$ channelized power values $P_k$ for each channel $k$ from the spectrometer. Values that deviate from unity beyond analytically determined thresholds are flagged. This is denoted with ${SK}_k$ and done by constructing two sums 
\[S_{1,k} = \sum_{m=1}^M P_k(m) \ \ \mbox{and} \ \  S_{2,k} = \sum_{m=1}^M P^2_k(m).\]  
The SK detection statistic,
\begin{equation}
{\rm SK}_k = \frac{M+1}{M-1} \left( \frac{MS_{2,k}}{S^2_{1,k}} -1 \right).
\label{eq:generalizedSK} 
\end{equation} 
Any data flagged outside of a threshold {which is often chosen to be}  $\pm 3 \sigma \approx \pm 6/\sqrt{M}$ on the ${SK}_k$ is considered RFI {with the threshold optimized for a given situation}. \citet{nita2007} have continued to improve upon this algorithm by generalizing it so the spectral averages may be taken before the SK estimator is calculated and using it in the two-bit digitized time domain \citep{nita2010, nita2016, nita2019, gary2010, Taylor2019}. 

\vskip \medskipamount
\noindent
{\bf Median absolute deviation (MAD)}
The MAD statistic for RFI detection in radio astronomy was proposed by Buch et al. \citep{Buch2016}. Its FPGA prototype was later developed by Ramey et al. \citep{Real-timeRFI:eramey} and by Buch et al. \citep{Buch2019}.  MAD uses the first-order statistic of the median to develop a decision rule to flag RFI. Its mathematical formulation is as follows. 

Let the median of data set $X$ be represented by 
\begin{equation}
M_{X} = {\rm median}(X),
\label{eq:medianData}
\end{equation}
and the median of the absolute deviation of the data set $X$ from $M_{X}$ be denoted by
\begin{equation}
\upsilon = {\rm median}(|X - M_{X}|).
\label{eq:MADdata}
\end{equation}
Given the two statistics $M_{X}$ and $\upsilon,$ the MAD decision rule is formed by comparing the absolute deviation of any given point within the set $X$ from the median $M_{X}$ with the threshold ${A\sigma_{r}}$, where $A$ is often chosen to be 3 but is optimized for particular situations. In general, we have
\begin{equation}
|x_{i} - M_{X}| \lessgtr A \sigma_{r},
\label{eq:MADdecisionRule}
\end{equation}
where $x_{i}$ is the $i$-th sample point of the data set and the robust standard deviation is
\begin{equation}
\sigma_{r} = 1.4826 \times \upsilon.
\label{eq:robustSTD}
\end{equation} 
Any sample outside the chosen deviation range is considered RFI. 

\subsection{Exploring Statistical Goodness-of-Fit Tests} 
\label{sec:goodness_of_fit_tests}
Since the Gaussian nature of RFI-free complex channelized voltage data is the main feature for distinguishing between the RFI-free data and the data containing RFI, involving normality tests developed to differentiate between Gaussian and non-Gaussian statistics would be a natural approach to the problem of RFI detection. One of the most popular tests for normality in statistics is the Shapiro-Wilks (SW) test \citep{Shapiro1965}.  In addition to its mathematical simplicity, it has the benefit of being easily parallelizable when implemented by hardware. It is for these practical reasons that SW is chosen over other similar tests such as the Anderson-Darling test.  

The idea behind the SW normality is simple and elegant. Given a set of samples from a standard Gaussian distribution and a query set of samples, each sorted in the order of increasing values and then plotted in pairs, if a straight line can be fitted to the pairs of sorted samples, then the query set is Gaussian in its nature. Otherwise, the Gaussian hypothesis is rejected. To describe the test mathematically, Shapiro and Wilks developed a dimensionless statistic by solving a generalized least-squares problem.  The developed statistic is described as  
\begin{equation} 
W = \frac{\sum_{i=1}^n a_i x_{(i)}}{\sum_{i=1}^n (x_i - \bar{x})^2}, 
\end{equation} 
where $x_i$ is the original unsorted $i$-th sample, $x_{(i)}$ is the $i$-th sorted sample, $\bar{x}$ is the sample mean. The coefficients $a_i$ form a vector 
\begin{equation}
(a_1,\ldots,a_n) = \frac{\mathbf{m}^T \mathbf{V}^{-1}}{(\mathbf{m}^T\mathbf{V}^{-1}\mathbf{V}^{-1}\mathbf{m})^{1/2}},
\end{equation} 
where $\mathbf{m}$ is the vector of sorted mean values of the samples from a standard Gaussian distribution and $\mathbf{V}$ is the covariance matrix of the sorted samples from the same standard Gaussian distribution.   Based on the statistical analysis performed by Shapiro and Wilks, the hypothesis that the set of query samples is Gaussian is accepted if the test's $p$-value  (the probability that the Gaussian distribution occurred by chance)  is larger than the $\alpha$-level (a preset conditional probability of error) of the test. The Gaussian hypothesis is rejected otherwise. 

\subsection{Exploring information theoretical performance metrics}
\label{sec:ITmetrics}
As a subject, information theory (IT) characterizes the achievable limits in designing efficient, high-performance communication systems \citep{Cover2006}.   Attributed to \citet{Shannon1948}, in the past 70 years, IT grew into a large discipline, overlapping with and in part encompassing both statistics and physics. As a result, the concept of entropy has a strong presence in both physics and IT. Entropy in physics was developed as a precise mathematical way of testing if the second law of thermodynamics holds in a particular process. Entropy in IT was developed to quantify the uncertainty (average self-information) in a random variable and the limit of lossless compression \citep{Shannon1948}.    Relative entropy also called the Kullback-Leibler divergence  was proposed as a metric to quantify the penalty for using the wrong probability distribution in lossless data encoding. It was later realized that it can be treated as a distance between two probability distributions \citep{Moulin2020}. Although relative entropy is not a real distance, since it does not satisfy the triangular inequality, it is a popular means to differentiate between two probability distributions in communication theory. 

Before introducing new IT-based statistics for testing the Gaussianity of channelized complex voltages, we formally define entropy and relative entropy. Given a probability mass function (pmf) $p(x)$ of a random variable $X$, its entropy is
\begin{equation}
H(X) = - \sum_x  p(x) \log p(x),
\label{eq:entropy}
\end{equation} 
i.e., the average negative logarithm of the probability.
The relative entropy, $D(p,q)$, between two pmfs $p(x)$ and $q(x),$ defined on the same set of outcomes, is the average of the log-likelihood ratio of $p(x)$ to $q(x)$, where the average is evaluated with respect to  $p(x)$. The relative entropy is therefore
\begin{equation} 
D(p,q) =  \sum_x  p(x) \log \frac{p(x)}{q(x)}.
\end{equation} 

Spectral entropy (SE) is a spectral tool developed for speech signal processing \citep{Shen1998RobustEE}. Unlike SK, which relies on the estimates of the mean and variance of pixel intensities in a spectral channel and on their ratio, SE is evaluated using the estimate of the probability mass function for each possible digitized voltage level, e.g., $2^8 = 256$ possible values of $x$ for 8-bit data. Similar to the SW test, SE relies on the assumption that the RFI statistic is not Gaussian. First, SE per channel is evaluated (following Eq.~\ref{eq:entropy}) together with the sample estimate of the digitized voltage variance per channel. Then the entropy of a Gaussian random variable  is evaluated  
\begin{equation}
H_{\rm base}(X) = \frac{1}{2} \left[1+ \log(2\pi\sigma^2)\right],
\end{equation}
where $\sigma^2$ is the estimated variance in an analyzed channel, and the entropy has units of nats, if we use the natural logarithm. The rest of the analysis relies on the fact that RFI-free channels have entropy close to the Gaussian entropy with the variance of the channel, and thus, the absolute difference in entropy values $|H(X) - H_{\rm base}(X)|$ is almost zero. 

\begin{figure}
\centering
\includegraphics[width=0.47\textwidth]{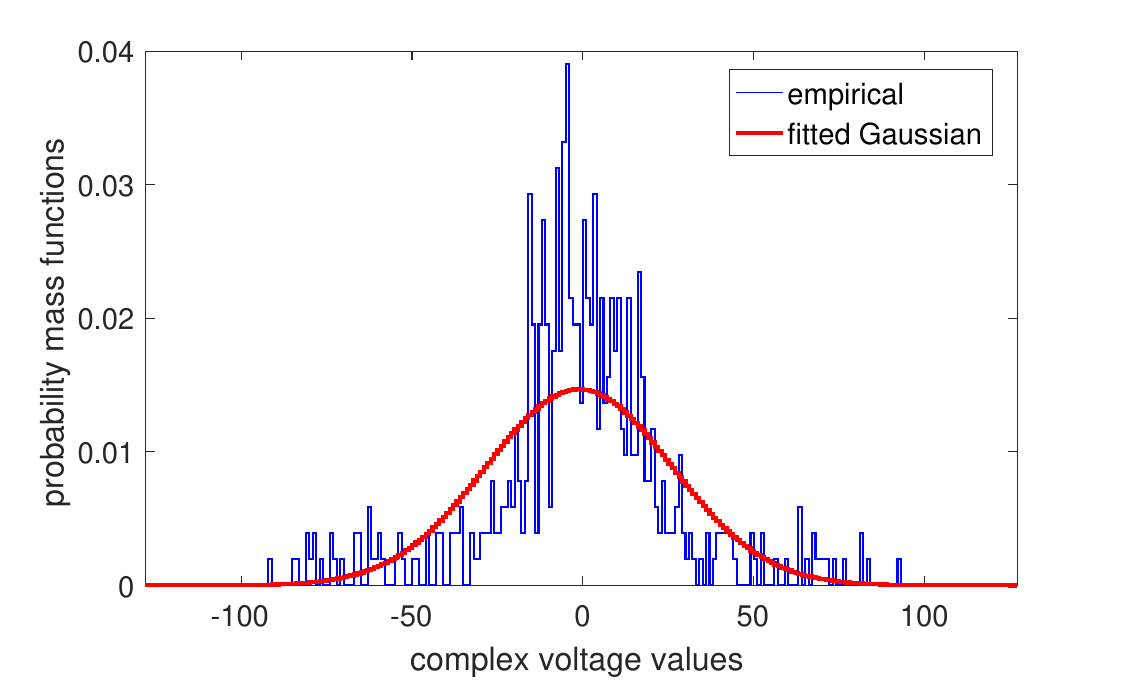}
\caption{The blue bar plot shows the empirical probability mass function of the real part of complex voltage values in channel 1830. The red line is the fitted Gaussian distribution with the mean and variance of the empirical data. The substantial deviation of the shape of the empirical probability mass function from the fitted Gaussian pdf is due to the presence of strong RFI in channel 1830. }
\label{fig:pmfs_channel_1830}
\end{figure}

To demonstrate the difference between an actual histogram of the time samples per frequency channel and a normalized fitted Gaussian curve with the mean and variance of the empirical data, both are displayed in Fig.~\ref{fig:pmfs_channel_1830}. The histogram has the number of bins based on the bit-resolution of the complex-valued channelized voltages. An 8-bit signed complex-valued channelized voltage would have $2^8 = 256$ bins ranging from --127 to 128. 

\begin{figure*}
\begin{center}
\begin{tabular}{cc}
    \includegraphics[width=0.45\textwidth]{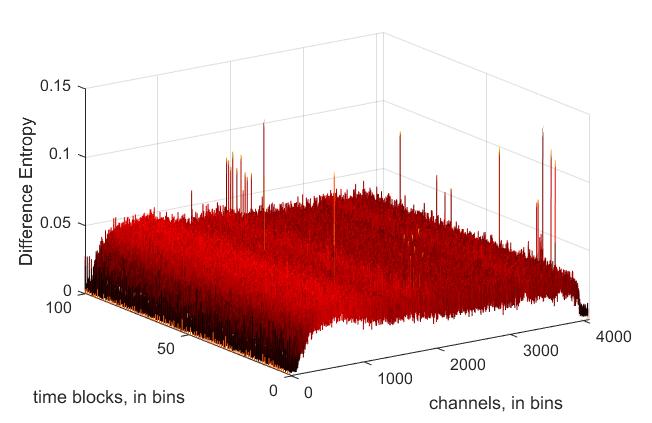} &
    \includegraphics[width=0.45\textwidth]{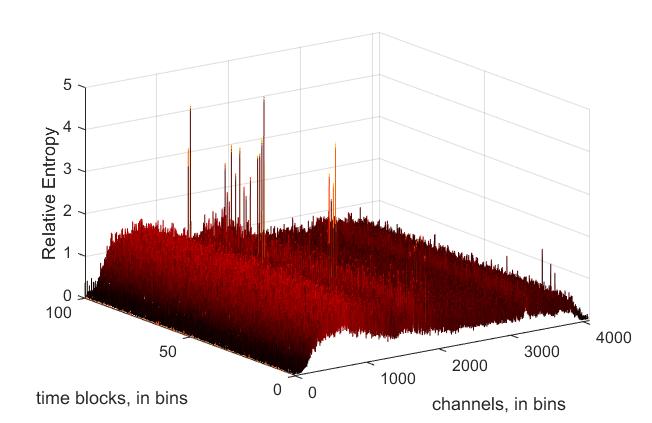}
\end{tabular}
\end{center} 
\caption{Left: Normalized spectral entropy as a normality test. Right: Spectral relative entropy as a test for normality.}
\label{fig:difference_entropy} 
\end{figure*}

To illustrate the potential of SE for the detection of RFI, Fig.~\ref{fig:difference_entropy} displays the absolute difference between the empirical and Gaussian SE values as a function of the number of spectral channels and time samples grouped by 512 original (high resolution) time samples. The original channelized voltage data block of size $65,024 \times 4096$ is partitioned into $127$ non-overlapping segments, each of size $512 \times 4096.$ The $512$ time samples per channel are used to calculate a single value of $|H(X) - H_{\rm base}(X)|.$ Note that the absolute difference SE can easily detect several sources of RFI present in the data. 

Finally, the RFI detection rule implements the modified $Z$-score, an efficient statistical method for detecting data outliers \citep{Iglewicz1993}, applied to the values of $|H(X) - H_{\rm base}(X)|.$  The modified $Z$-score is the same outlier detection method as in the MAD algorithm (see (\ref{eq:medianData}) through (\ref{eq:robustSTD})). The only difference is that the data points in the MAD rule are replaced with the values of $|H(X) - H_{\rm base}(X)|.$ This choice of the decision rule ensures that channels are treated independently, enabling a parallel implementation on GPU. 

Spectral relative entropy (SRE) yields a powerful test for Gaussianity, provided that the reference distribution $p(x)$ is selected to be normalized Gaussian with the mean and variance estimated per channel from empirical data.  Unlike SE, which gains its power due to the subtraction of the Gaussian entropy, 
 SRE relies not only on the difference in shapes of the two involved pmfs but also on the difference in terms of their higher-order statistics.   

Similar to the computation of SE, we first find the estimate of the pmf of the channelized quantized voltages (in our example, time segments are grouped in sets of 512 original samples).  As a second step, we evaluate the sample mean and sample variance per channel per segment.  Next, we fit a Gaussian probability density function with the sample mean and variance of the data in the channel minimizing the least-squares metric. Since the empirical pmf is based on an 8-bit representation, the Gaussian pdf is sampled at 256 locations of the empirical pmf bin centers.  Finally, the relative entropy between the fitted Gaussian and the empirical pmf of channelized voltage levels (over a given time segment) is evaluated. The right panel in Fig.~\ref{fig:difference_entropy} displays the plot of SRE as a function of the number of spectral channels and time samples grouped by 512 original (high resolution) time samples. Note that the information in this plot is much more refined than the information provided in the plot of the normalized spectral entropy.  This difference is attributed to the fact that the relative entropy measure contains information not only about the shape of two individual probability density functions but also about other high-order statistics describing the data. 

Similar to the case of SE, the detection rule implements the modified $Z$-score but is applied to SRE values, {with a $Z$-score magnitude greater than 3 often chosen as a threshold, but is optimized for a given environment}. In addition to the relative entropy between theoretical and estimated pmfs, we also look at the symmetrical case of SRE formed by the summation of two asymmetrical SREs: 
\begin{equation} 
\begin{split}
D_{\rm sym}(p,q) & = D(p,q) + D(q,p) \\ 
& = \sum_{x}p(x)\log{\frac{p(x)}{q(x)}} + \sum_{x}q(x)\log{\frac{q(x)}{p(x)}}.
\end{split}
\label{eq:RelEntropy_sym}      
\end{equation} 
To differentiate between the two cases of SRE, we name the symmetrical case as SRE$_{\rm s}$ and the asymmetrical case as SRE$_{\rm a}$.

\section{Data} 
\label{sec:data}
In this section, we describe the characteristics of the data that RFI detection tests were performed on. All data are illustrated with the higher frequency channel as the lowest frequency in the bandwidth and the lowest frequency channel as the highest frequency in the bandwidth. To be more specific, the data in use were collected over a bandwidth of 800~MHz partitioned into 4096 non-overlapping frequency channels. Channel 4096 corresponds to 1100~MHz, whereas channel 1 corresponds to 1900~MHz. The data are in the form of high-resolution complex-valued channelized voltages at two polarizations, polarization 0 and polarization 1. 
To calculate the power spectral density (PSD) from the complex-valued channelized voltages, real and imaginary parts at each discrete time and frequency location are squared and summed. 
\begin{table}
        \centering
        \begin{tabular}{|c|c|}
            {PSR} & {J1713+0747} \\
            \hline \hline
            {Sampling Interval ($\mu$s)} & {5.12} \\
            {Length of Data (s)} & {1.6646144} \\
            {Number of Time Samples}  & {325,120} \\
            {Number of Frequency Channels}  & {4096} \\
            {Bandwidth (MHz)}  & {800} \\
            {Center Frequency (MHz)} & {1500} \\
            {Number of Bits} & {8} {signed} \\
            \hline          
        \end{tabular}
        \caption{This table describes the characteristics of the data set containing astronomical pulses,  RFI, and noise. }
        \label{table:pulseFreeData}
\end{table}
%
%

The aforementioned RFI detection methods are tested on observations containing both RFI and pulsar signals. The pulsar in question, PSR~J1713+0747, is a millisecond pulsar with period $P= 4.5$~ms, a pulse width of 1~ms, and a DM of 15.97~cm$^{-3}$~pc \citep{1993ApJ...410L..91F}. These data also contain known RFI from: (i) the Iridium satellite communication system over the frequency range 1620-1626 MHz (channels 1402--1433); (ii) FAA radar originating from the Bedford, NC station is seen at 1255~MHz and 1305~MHz (channels 3302 and 3046); (iii) a collection of unknown sources that exists around 1500~MHz (channel 2048). Periodic RFI from GPS-L3 Communications is found at 1381~MHz (channel 2657). 

The data were collected by our colleagues at GBO on eight different occasions and thus spaced in time and saved as eight distinct data files each of size 5~GB. The raw complex voltages saved in each file were sampled at a Nyquist frequency of 1600 MHz, then converted to channelized voltages by means of the short-time FT. The complex channelized data files are available to us as real and imaginary parts each of size 4096 frequency channels, 325,120 time samples, and two polarizations. After channelization, the sampling interval is reduced to 5.12~$\mu$s. There are about $1.66$~s of test data in each file. This means that the number of time samples representing a complete period of the pulsar is about 892 and the maximum number of pulses that can be found in the test data is approximately 370. The right column of Table \ref{table:pulseFreeData} summarizes the parameters of the dataset. 

Each file was shared with us in Matlab data file format and was saved in 5 non-overlapping chunks of size 65,024-by-4096 at a time resulting in a phase discontinuity of the pulsar pulse at the end of each 65,024-th time sample. Therefore, to avoid any misleading results, we partitioned each file into 5 chunks, each containing 65,024 time samples and 4096 frequency channels. 
To distinguish between files and chunks, we named each file as $mat \ number,$ with numbers ranging between 0 and 7, and each chunk as $chunk \ number,$ with numbers ranging between 0 and 4.

Each data chunk was further broken down into non-overlapping, consecutive segments containing all of the frequency channels and 512 time samples. Thus, 127 segments of 512 time samples and 4096 frequency channels were formed per each $chunk \ number$ file.

The MAD algorithm, SW test for normality, SK, SRE, and SE are applied to each data segment. Since there is no way to definitively know what and where RFI signals are, there is no definitive ground truth.  To analyze the effectiveness of the newly proposed methods in the detection and mitigation of RFI, we compare their performance on the pulsar signal to noise against the performance of the MAD and SK methods, both known in the literature, on the same metric.

\begin{figure}
\begin{center}
\begin{tabular}{c}
    \includegraphics[width=0.45\textwidth]{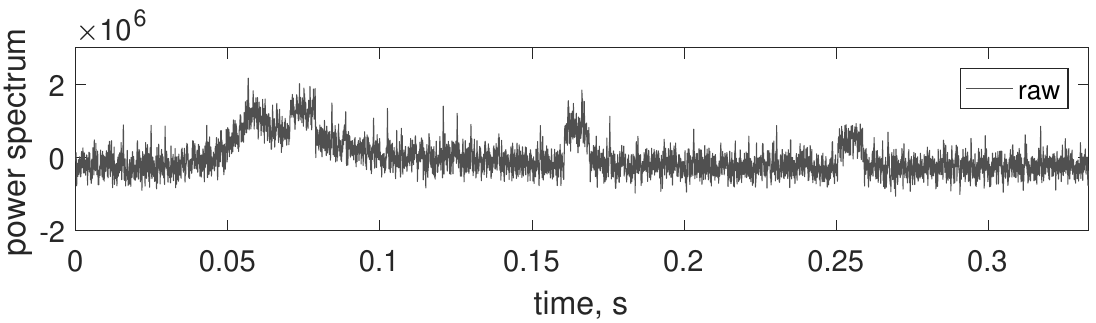} \\
    \includegraphics[width=0.45\textwidth]{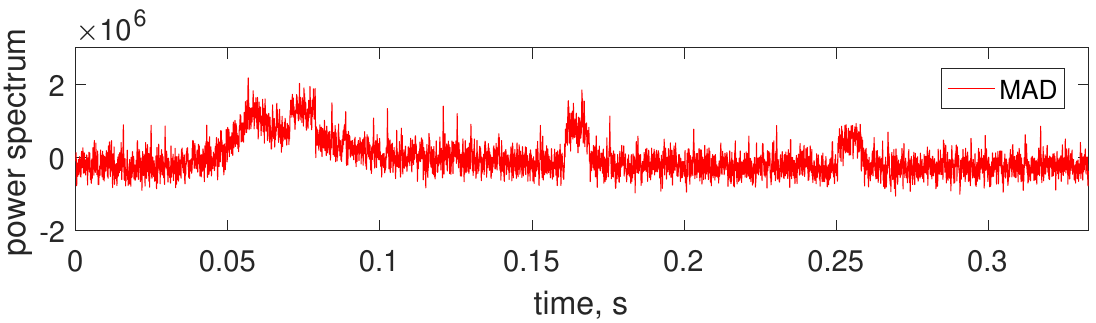} \\ 
    \includegraphics[width=0.45\textwidth]{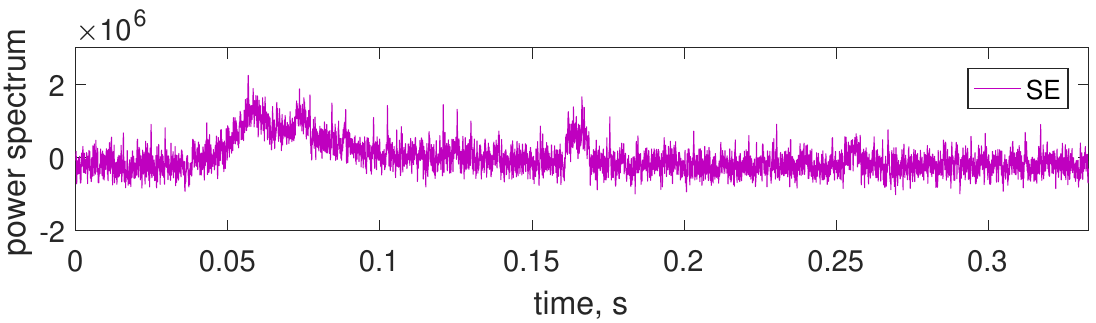} \\
    \includegraphics[width=0.45\textwidth]{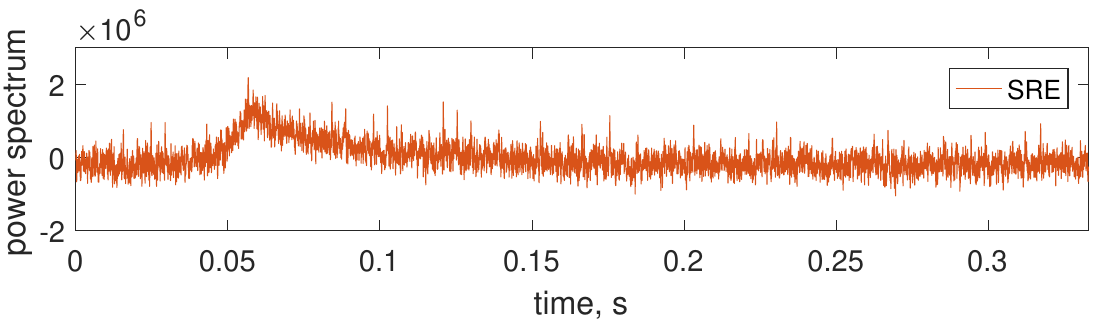} \\
    \includegraphics[width=0.45\textwidth]{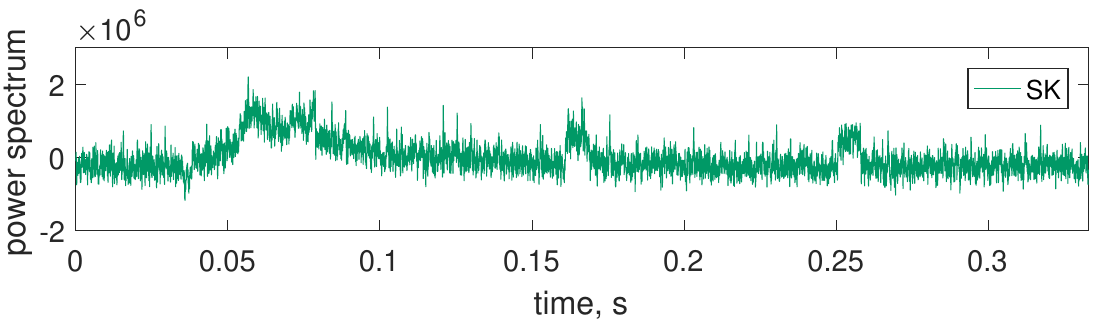} \\
    \includegraphics[width=0.45\textwidth]{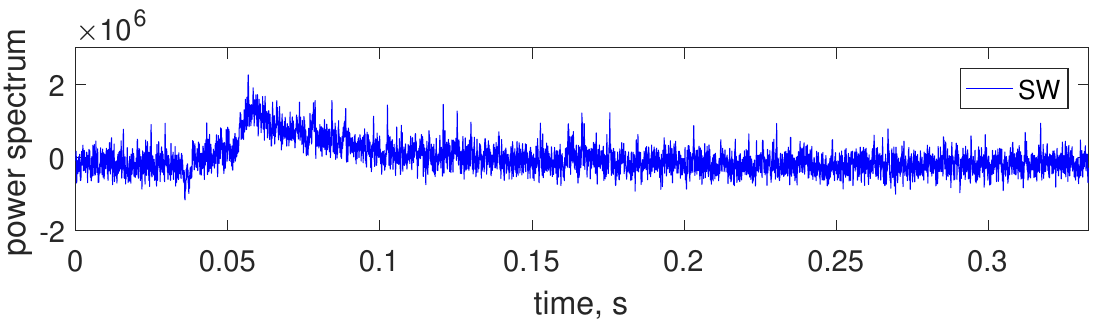}    
\end{tabular}
\end{center}
\caption{Power spectrum displayed as a time series. The time series contains 57 pulsar pulses embedded in noise and the remaining RFI.}
\label{fig:performance}
\end{figure}

To see a pulse in the data of J1713+0747, several processing steps must be applied. First, the channelized voltages must be converted to power spectral density by summing the squares of both the real and imaginary valued channelized voltages. Next, the data must be dedispersed using the DM value of the pulsar. Subsequently, integration of the dedispersed data over frequency components is completed.  A depiction of a single chunk of $mat \ 0$ after the application of the signal processing steps is shown in Fig.~\ref{fig:performance}. Several pulses of the pulsar are clearly seen in each panel between 0.1 and 0.15~s. 

\section{Experimental Results} 
\label{sec:results}
Given raw channelized voltage data as described in Section \ref{sec:data} 
and a list of prospective RFI detection and mitigation methods applied to the raw data, we now illustrate the performance of the RFI detection and mitigation methods. We adopt S/N as an objective measure of performance. The S/N of a single folded pulse is a traditional metric to measure the quality of astronomical signals when searching for pulsars \citep{Lorimer12}.

After the inspection of the eight data files, we selected two files, $mat \ 0$ and $mat \ 2$, due to the unique types of RFI present in the data. The first file $mat \ 0$ contains several broadband RFI signals, while $mat \ 2$ has a presence of strong RFI signals varying in frequency. Both types of RFI present challenges for modern RFI detection methods. 
Fig.~\ref{fig:performance} and Tables \ref{table:comparison_chunk_0}--\ref{table:comparison_chunk_4} display the results of our analysis of five different RFI detection methods defined in Sections \ref{sec:other_methods}, \ref{sec:goodness_of_fit_tests}, and \ref{sec:ITmetrics} in application to the five chunks of $mat \ 0$. Tables \ref{table:comparison_mat_2_chunk_0}--\ref{table:comparison_mat_2_chunk_4} demonstrate the results of our analysis in application to the five chunks of $mat \ 2.$  

\subsection{Performance Analysis of $mat \ 0$}
\begin{figure}
\centering
\includegraphics[width=0.45\textwidth]{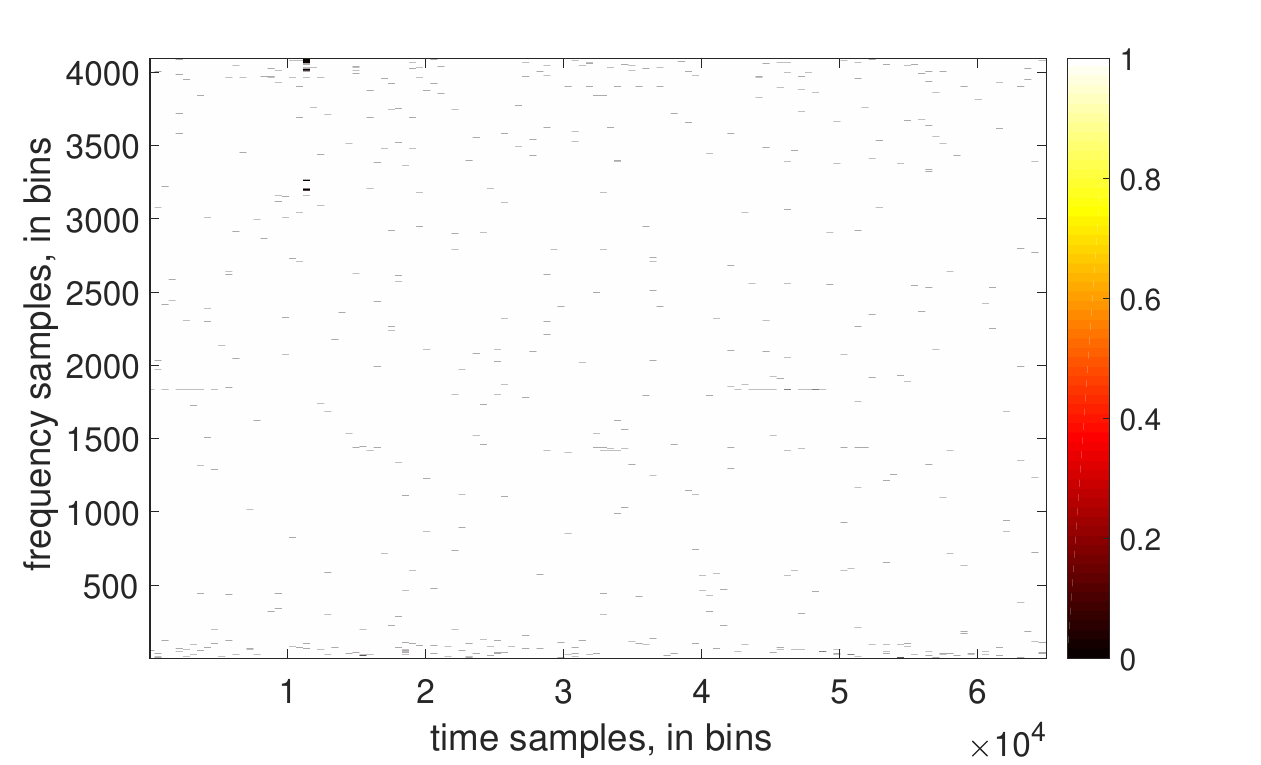} 
\caption{Mask generated by SE at the value of threshold set to $4\sigma.$ 
Small black intervals mark detected RFI, where the test rejected the Gaussian hypothesis. White intervals mark the part of the data where the test did not reject the Gaussian hypothesis.}
\label{fig:diffE_fb8}
\end{figure}

\begin{figure}
\begin{center}    
\begin{tabular}{c}
\includegraphics[width=0.45\textwidth]{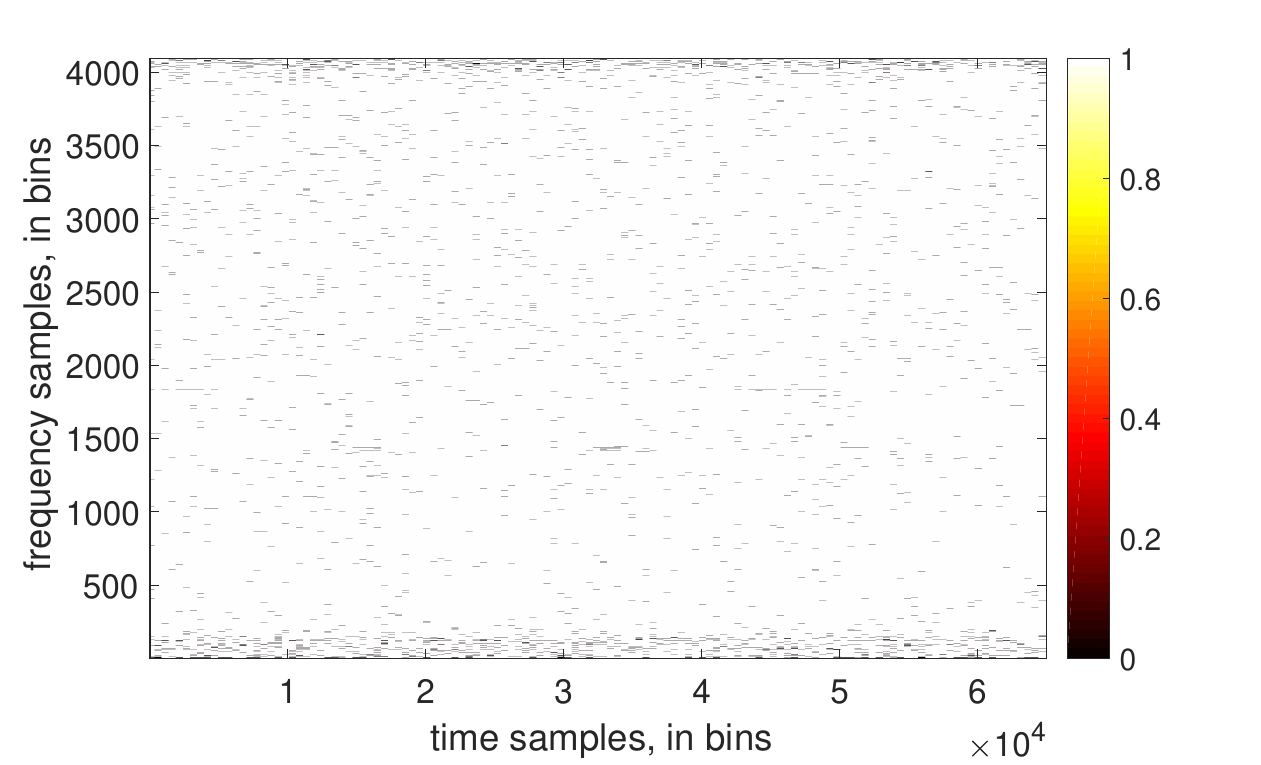} 
\end{tabular}
\caption{Mask generated by asymmetrical SRE at the threshold of $4\sigma.$ For further details, see Fig.~7}
\label{fig:KL_mask2}
\end{center}
\end{figure}

\begin{figure}
\begin{center}    
\begin{tabular}{c}
\includegraphics[width=0.45\textwidth]{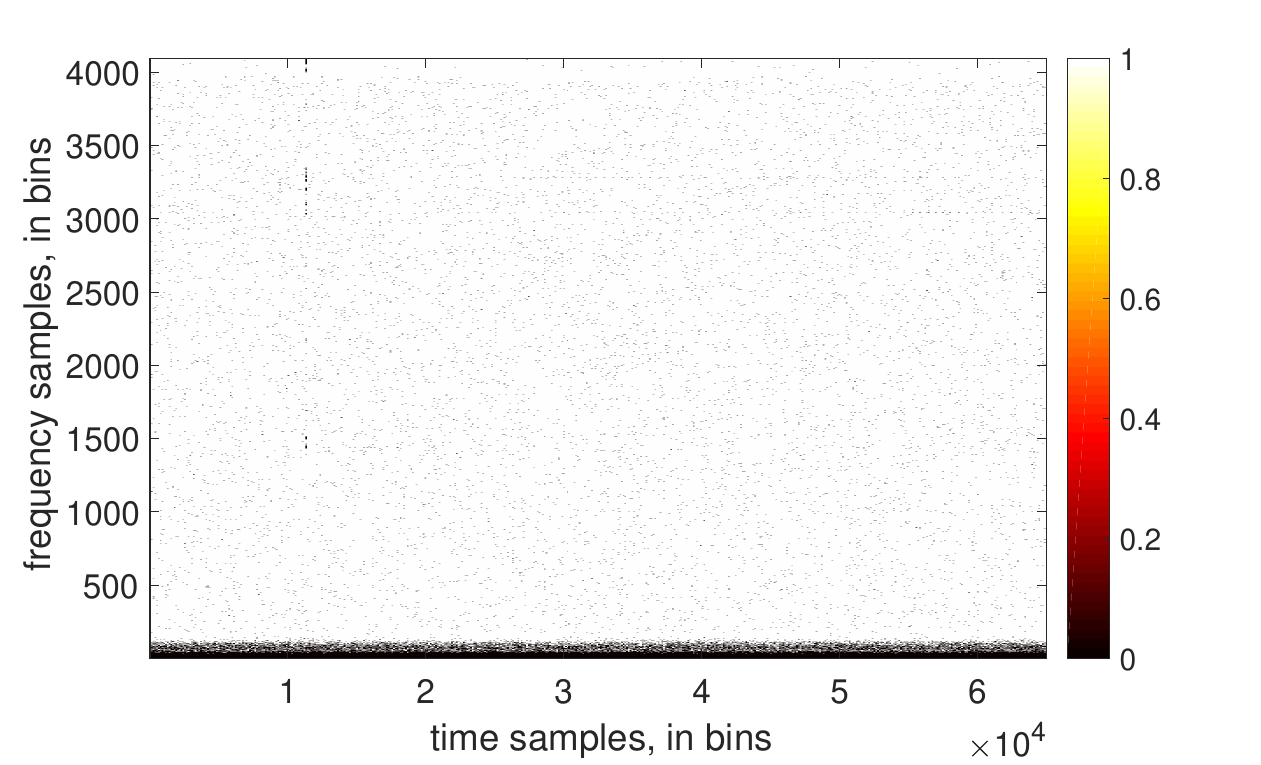} 
\end{tabular}
\caption{Mask generated by SK at the threshold of $3\sigma.$ For further details, see Fig.~7}
\label{fig:KL_mask3}
\end{center}
\end{figure}

\begin{figure}
\centering
\includegraphics[width=0.45\textwidth]{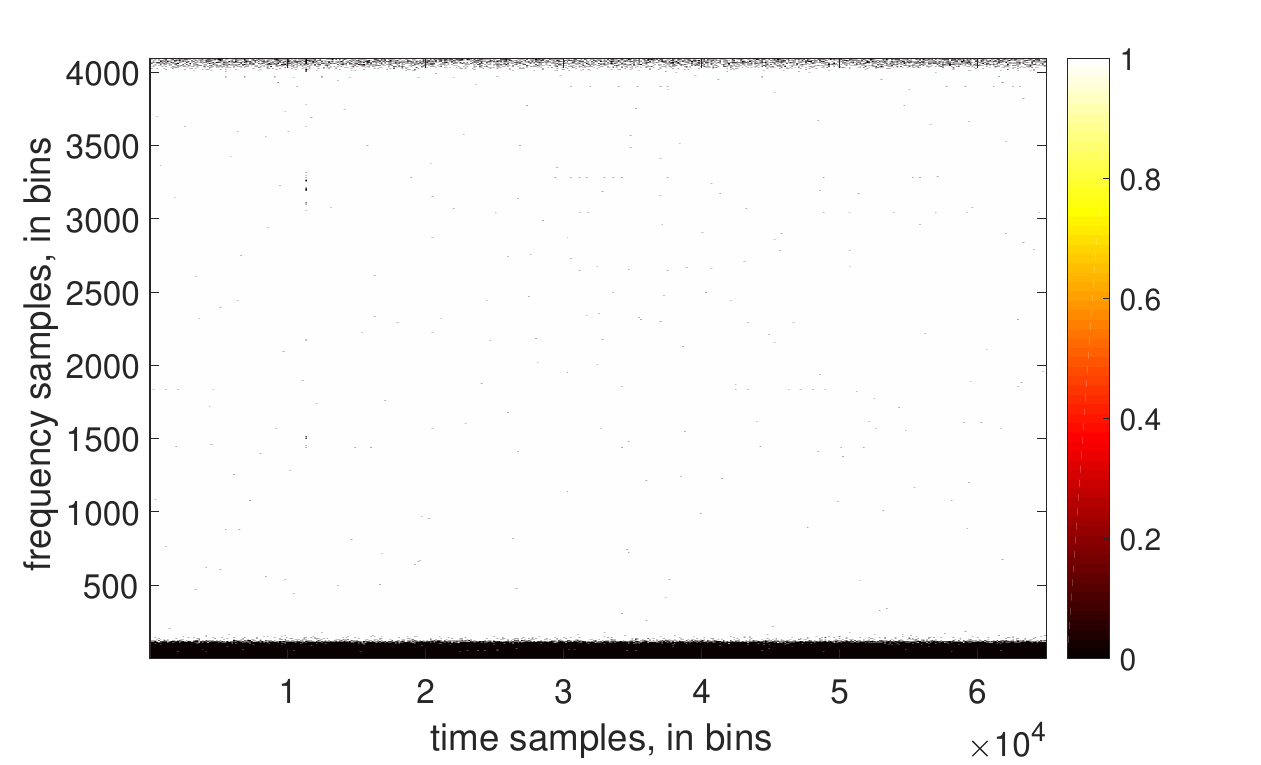}
\caption{Mask generated by SW using an $\alpha$-level set to $10^{-4}$. For further details, see Fig.~7.}
\label{fig:SW_mask}
\end{figure}

As mentioned earlier, the data file $mat \ 0$ was selected for analysis due to its unique content. The file contains several broadband RFI signals. One of them is shown in the form of ``RFI masks'' in Figs.~\ref{fig:diffE_fb8}--\ref{fig:SW_mask}.  Since complex channelized voltage data are represented by real and imaginary parts, a mask is generated per each part, then a single combined mask is generated as a product of the two masks. Different RFI detection methods are applied to $chunk \ 0$ of $mat \ 0$ yielding several combined masks, one per each method.  The chunk is of size 65,024 time samples and 4096 frequency channels. It is partitioned into 127 non-overlapping segments, each composed of 512 time samples and 4096 channels. The RFI detection methods are applied to 512 time samples in every channel and every segment. If a particular test detects the presence of RFI, the 512 time samples are replaced with zeros, otherwise, they are replaced with ones. Black lines and bars mark detected RFI while white space represents the portion of the data free of RFI as determined by each RFI detection method. The illustrations are provided for zero polarization of the data. The RFI masks are shown for SE at the threshold of $4\sigma,$ for SRE at the threshold of $4\sigma,$ for SK at values of the threshold of $3\sigma,$ and for SW at the $\alpha$-level of $10^{-4}.$ Note that although SK and SW were applied for the detection of narrow-band RFI (along each frequency channel), they also captured broad-band RFI, unlike MAD, SE, and SRE methods.  We do not show the RFI mask for MAD since, regardless of the threshold, it does not display any essential RFI signals. 
%
%
%
%
%
%
%

After RFI masks are generated, several signal processing steps are applied to the data to arrive at the plots in Fig.~\ref{fig:performance}. The steps are: (1) applying the combined RFI masks to real and imaginary parts of complex-valued channelized voltages (multiplying them one-by-one); (2) forming the power spectrum (spectrogram); (3) dedispersing the data; (4) integrating dedispersed data in frequency. The outcome is a power spectral time series.  

Six integrated power spectral series are displayed in Fig.~\ref{fig:performance}. The top panel shows the raw power spectrum. The second panel shows the power spectrum after the MAD method at $5.5\sigma$ was applied. The third from the top panel presents the power spectrum after applying SE at $4\sigma.$ The fourth panel shows the power spectrum after the application of SRE at $5\sigma.$ The fifth panel displays the power spectrum after applying SK at $6.5\sigma$ and the panel at the bottom shows the power spectrum after applying the SW test with $\alpha$ set to $10^{-4}$. Note that the thresholds were selected to maximize the performance of each detection method as will be explained below. 

To quantify the performance of the proposed RFI detection methods, we compute the S/N values of a folded pulse for different methods. The results are summarized in Table \ref{table:comparison_chunk_0}. To arrive at each S/N, a single folded pulse is generated from the power spectrum shown in Fig.~\ref{fig:performance} using {\sc riptide} \citep{Morello_2020}, a Python implementation of the fast folding algorithm \citep{1969Fast_Folding_Algorithm}. Table~\ref{table:comparison_chunk_0} displays the found S/N values as a function of varying thresholds and $\alpha$-levels. Thresholds for the methods of MAD, SE, SRE, and SK are varied between $3\sigma$ and $7\sigma.$ The values of $\alpha$-level used by SW are varied between 0.01 and $10^{-5}$.  

\begin{figure}
\centering
\includegraphics[width=0.45\textwidth]{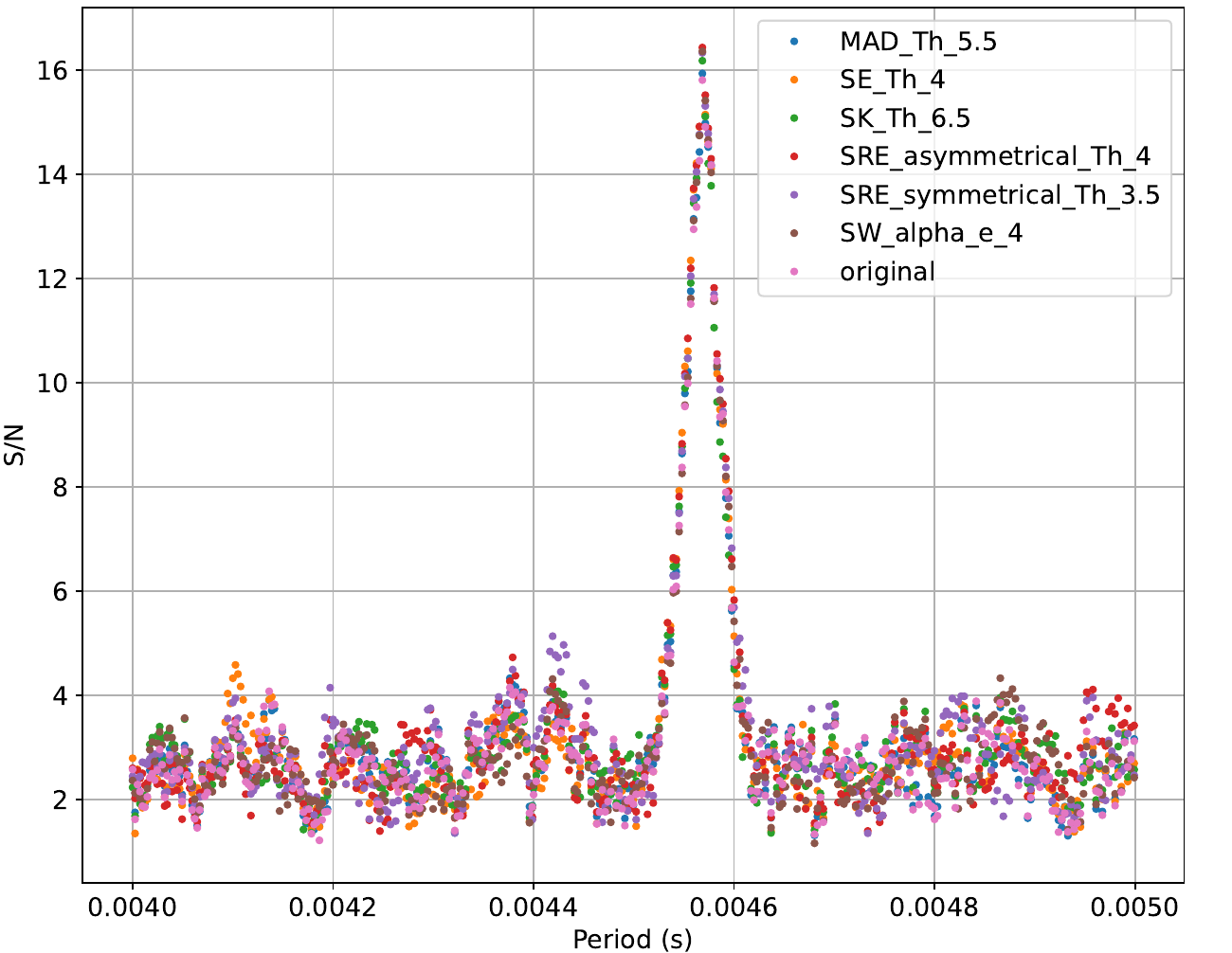} 
\caption{Periodogram results from {\sc riptide} for the proposed and baseline RFI detection methods obtained from the data in $chunk \ 0$ of $mat \ 0.$}
\label{fig:chun0_mat0_SN}
\end{figure}

Table \ref{table:comparison_chunk_0} provides insight into the best performance delivered by each RFI detection method, given the data partitioning as described in Section \ref{sec:data}.  
MAD, one of the two baseline methods selected for performance comparison, achieves the maximum S/N value of $15.94$ at the threshold of $5.5\sigma.$ This is slightly above the untreated (raw data) S/N of $15.81.$ When the threshold is set to $3\sigma,$ as recommended in \citep{Real-timeRFI:eramey}, the S/N of MAD is below the S/N of raw (untreated) data. Each of the remaining four methods (SE, symmetrical SRE, asymmetrical SRE, SK, and SW), demonstrates a more significant performance improvement compared to MAD. As an example, SE achieves the best performance of $16.35$ at $4\sigma,$ symmetrical SRE achieves S/N of $16.33$ at the threshold $3.5\sigma,$ asymmetrical SRE achieves S/N of $16.43$ at the threshold $4\sigma,$  SK reaches S/N of $16.18$ at the threshold $6.5\sigma,$ and SW demonstrates the S/N value of $16.36$ at $\alpha$-level set to $10^{-4}$. {To conclude, our proposed methods are all better than the baseline methods of MAD and SK in the case of $mat \ 0 chunk \ 0$. It should be also noted that asymmetrical SRE is performing better than symmetrical SRE.} The plots of a single folded pulse for the choice of the best S/N value for the six RFI detection methods as well as for the original case are provided in Fig.~\ref{fig:chun0_mat0_SN}.  

\begin{figure}
\centering
\includegraphics[width=0.45\textwidth]{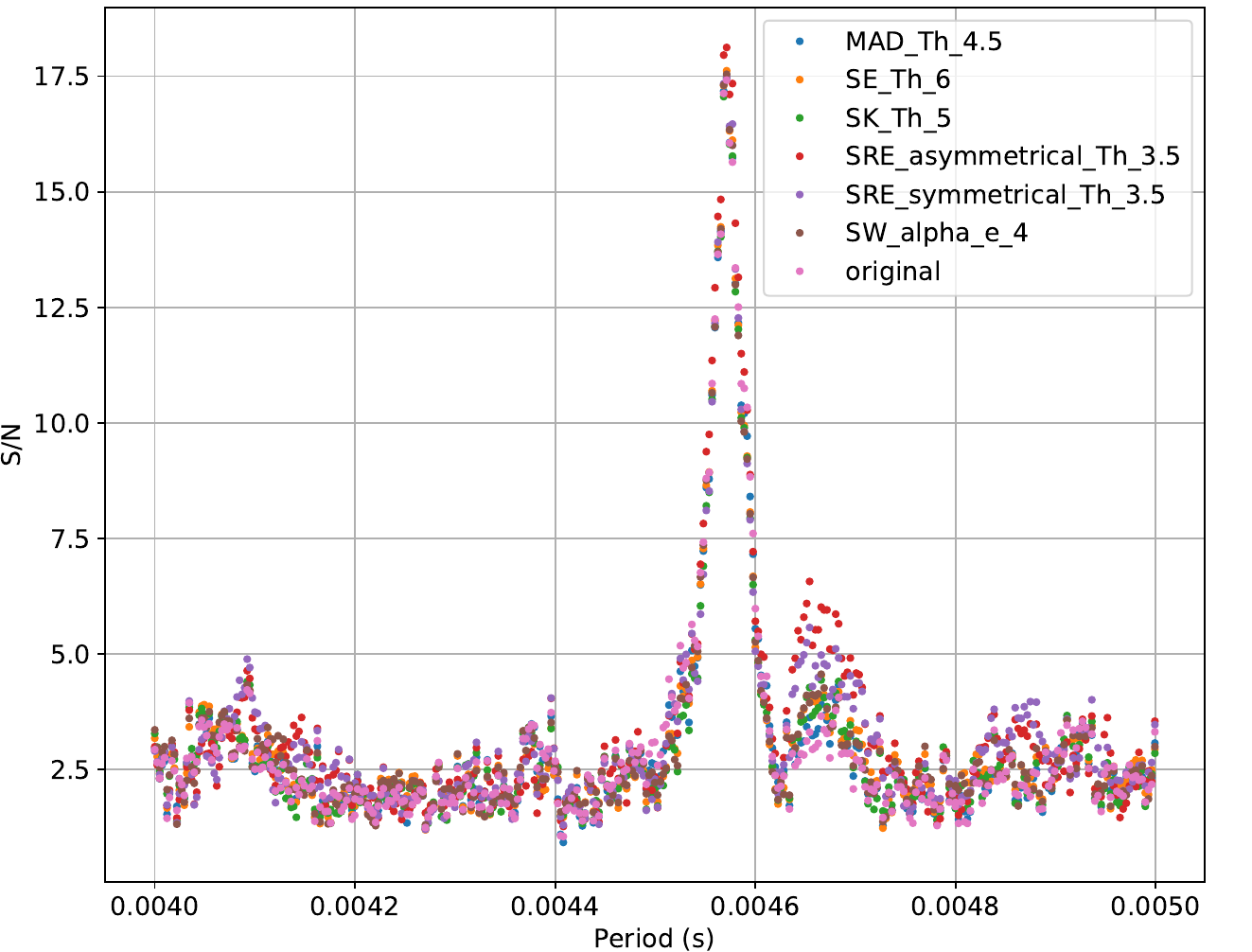} 
\caption{Periodogram results from {\sc riptide} for the proposed and baseline RFI detection methods obtained from the data in $chunk \ 1$ of $mat \ 0$.}
\label{fig:chun1_mat0_SN}
\end{figure}

To complete the analysis of $mat \ 0,$ we process the data in $chunks \ 1$ through 4.  The S/N value of raw data in $chunk \ 1$ of $mat \ 0$ is higher than the S/N of any other chunk. It is equal to $17.42$ for $chunk \ 1.$ Looking at the values provided in Table \ref{table:comparison_chunk_1}, there is no S/N value that is significantly higher than $17.42,$ pointing to the fact that the removal of RFI signals in this case is not that useful. Nonetheless, our proposed methods of SE, symmetrical SRE, asymmetrical SRE, and SW all surpass the baseline methods of MAD and SK. The highest of the best S/N equal to $17.62$ is achieved by SE with the threshold value set to $6\sigma,$ while the lowest of the best S/N equal to $17.45$ is achieved by MAD with the threshold value set to $4.5\sigma.$ The respective single pulse plot for $chunk \ 1$ of $mat \ 0$ is provided in Fig.~\ref{fig:chun1_mat0_SN}.

\begin{figure}
\centering
\includegraphics[width=0.45\textwidth]{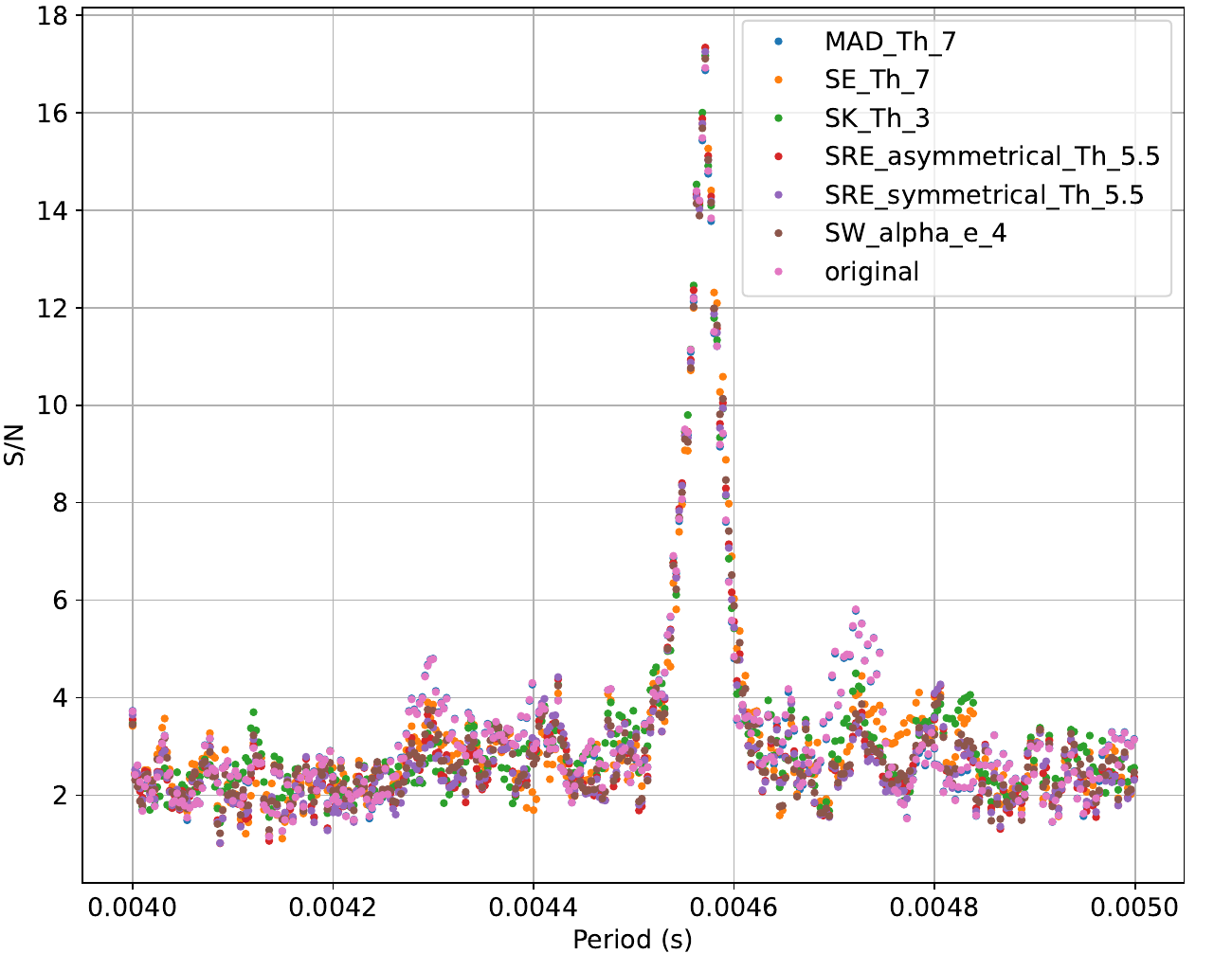} 
\caption{Periodogram results from {\sc riptide} for the proposed and baseline RFI detection methods obtained from the data in $chunk \ 2$ of $mat \ 0$.}
\label{fig:chun2_mat0_SN}
\end{figure}

\begin{figure}
\centering
\includegraphics[width=0.45\textwidth]{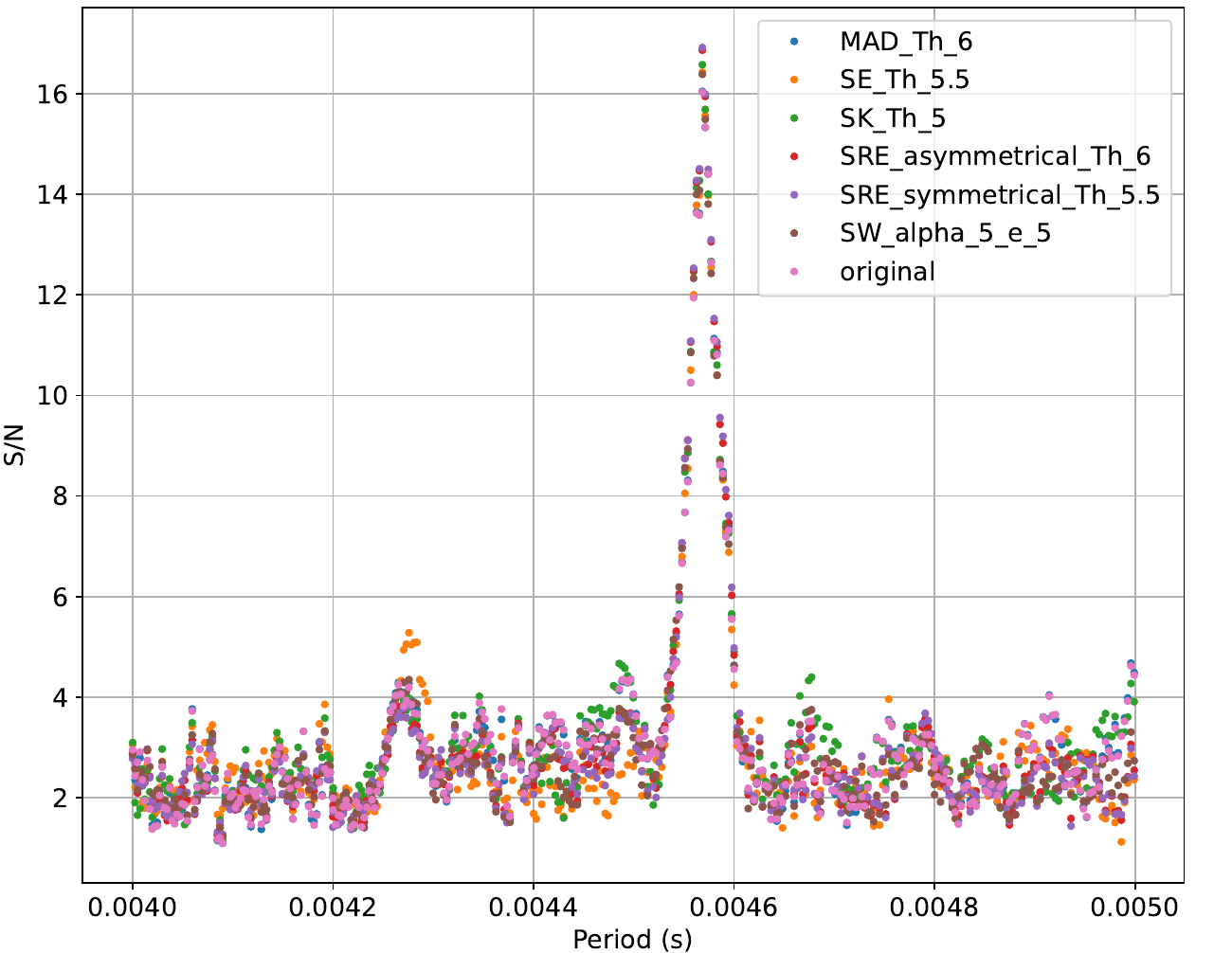} 
\caption{Periodogram results from {\sc riptide} for the proposed and baseline RFI detection methods obtained from the data in $chunk \ 3$ of $mat \ 0$.}
\label{fig:chun3_mat0_SN}
\end{figure}

\begin{figure}
\centering
\includegraphics[width=0.45\textwidth]{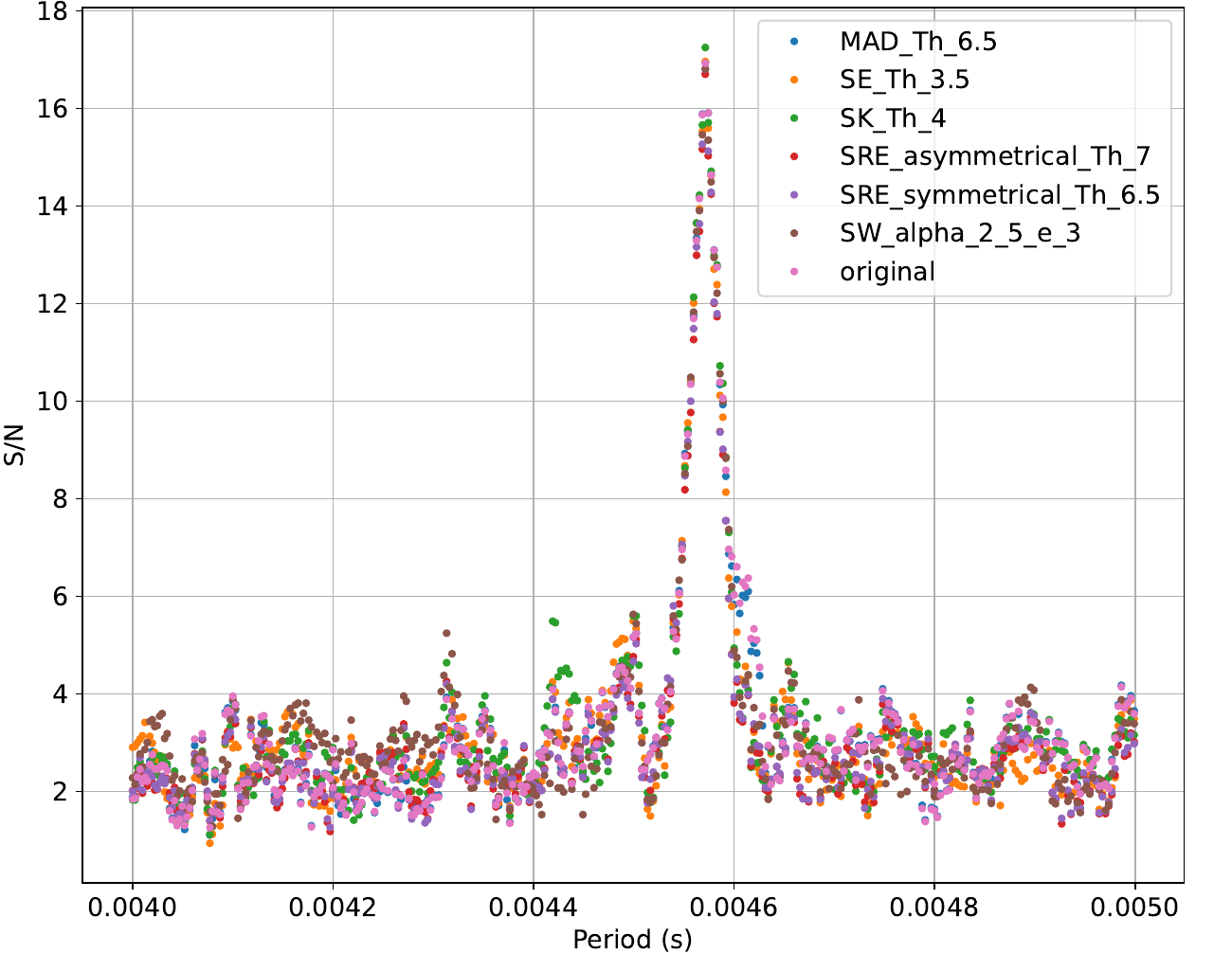} 
\caption{Periodogram results from {\sc riptide} for the proposed and baseline RFI detection methods obtained from the data in $chunk \ 4$ of $mat \ 0$.}
\label{fig:chun4_mat0_SN}
\end{figure}

The improvement in S/N is much more noticeable when the S/N value of raw data is relatively low. For example, for $chunk \ 2$ (see Table \ref{table:comparison_chunk_2}) the S/N value of raw data is $13.93.$ The application of SK at a threshold of $6.5 \sigma$ and SW at an $\alpha$-level of $10^{-4}$ result in S/N values of $17.66$ and $17.71,$ respectively, indicating that at a low value of S/N of the raw data, it is beneficial to detect and remove high in value RFI signals. The analysis of the S/N values as a function of the method of removal of RFI signals and varying threshold value ($\alpha$-level for SW) in Tables \ref{table:comparison_chunk_3} and \ref{table:comparison_chunk_4} demonstrate a similar trend.  The respective periodograms for the best values of S/N are shown in Figures \ref{fig:chun1_mat0_SN}-\ref{fig:chun4_mat0_SN}.

\begin{table}
\centering 
\begin{tabular}{ cccccccc } 
\toprule[1pt] 
 \multicolumn{8}{c}{ chunk 0 of mat 0 } \\
  \midrule[0.8pt] 
 Th  & MAD   & SE   & SRE$_{s}$  & SRE$_{a}$ & SK    & SW    & $\alpha$-level  \\ 
 \midrule[0.8pt] \midrule[0.8pt] 
 3   & 13.65 & 15.21 & 15.96 & 15.88 & 15.05  & 15.52 & 0.01            \\ 
 3.5 & 15.03 & 15.95 & \textbf{16.33} & 16.11   & 15.49 & 15.50 & 0.005           \\ 
 4   & 15.78 & \textbf{16.35}  & 16.14  & \textbf{16.43}  & 15.60  & 15.64 & 0.0025           \\ 
 4.5 & 15.82 & 16.30 & 16.29 &  16.38 &  15.83 & 16.18 & 0.001         \\ 
 5   & 15.93 & 16.29 & 16.30    &  16.34   & 15.90  & 16.35 & 0.0005        \\ 
 5.5 & \textbf{15.94}  & 16.23 & 16.21 & 16.20   &  16.17 & 16.33  & 0.00025   \\ 
 6   & 15.89 & 16.20 & 16.23 & 16.20  & 15.97 & \textbf{16.36} & 0.0001  \\ 
 6.5 & 15.87 & 16.15 & 16.28 & 16.17  & \textbf{16.18}  &16.12 &0.00005 \\ 
 7   & 15.85 & 16.07 & 16.23 & 16.20  &  16.18  &16.13 & 0.00001\\ 
 \bottomrule[1pt]
\end{tabular}
\caption{The S/N values on chunk 0 of mat 0 for different thresholds and different RFI removal methods of Median Absolute Deviation (MAD), Spectral Entropy (SE), Symmetrical Spectral Relative Entropy (SRE${_s}$), Asymmetrical Spectral Relative Entropy (SRE$_{a}$), Spectral Kurtosis (SK), and Shapiro-Wilks (SW). The maximum value of each method is marked in bold. The S/N of the raw data is 15.81.}
\label{table:comparison_chunk_0}
\end{table}

\begin{table} 
\centering 
\begin{tabular}{ cccccccc } 
\toprule[1pt] 
 \multicolumn{8}{c}{ chunk 1 of mat 0}  \\
  \midrule[0.8pt] 
 Th  & MAD   & SE   & SRE$_{s}$ & SRE$_{a}$ & SK    & SW    & $\alpha$-level  \\ 
\midrule[0.8pt]    \midrule[0.8pt]
 3   & 15.68 & 17.50 & 17.29 &  17.78  & 17.42 & 16.72 & 0.01            \\ 
 3.5 & 16.99 & 16.88 &  \textbf{17.48} & \textbf{18.12}  & 17.33 & 17.31 & 0.005           \\ 
 4   & 17.40 &  17.06 & 17.35& 17.55  & 17.26 & 17.32 &  0.0025         \\ 
 4.5 & \textbf{17.45}  & 17.22 &  17.34& 17.04  & 17.10 & 17.05  & 0.001           \\ 
 5   & 17.41 & 17.28 & 17.33& 17.15  & \textbf{17.47} & 17.26 & 0.0005           \\ 
 5.5 & 17.42 & 17.47 & 17.23&  17.33  & 17.42 & 17.45  & 0.00025 \\ 
 6   & 17.42 & \textbf{17.62}  &  17.31& 17.33  & 17.40 &\textbf{17.54} & 0.0001 \\ 
 6.5 & 17.43 & 17.49 & 17.25& 17.31  & 17.36  & 17.49 & 0.00005 \\ 
 7   & 17.42 & 17.43 & 17.22&  17.28 & 17.44  &17.39 & 0.00001 \\ 
 \bottomrule[1pt]
\end{tabular}
\caption{The S/N values on chunk 1 of mat 0 for different thresholds and different RFI removal methods. The S/N of the raw data is 17.42. } 
\label{table:comparison_chunk_1}
\end{table}

\begin{table} 
\centering 
\begin{tabular}{ cccccccc } 
\toprule[1pt] 
 \multicolumn{8}{c}{ chunk 2 of mat 0 } \\
  \midrule[0.8pt] 
 Th  & MAD   & SE   & SRE$_{s}$ & SRE$_{a}$ & SK    & SW    & $\alpha$-level  \\ 
 \midrule[0.8pt]  \midrule[0.8pt] 
 3   & 13.53 & 16.12 & 15.78& 16.22  & \textbf{17.17}  & 15.66 & 0.01            \\ 
 3.5 & 15.33 & 16.73 &  15.81& 16.30  & 17.13 & 16.18 & 0.005           \\ 
 4   &  16.17 & 16.92  & 16.55 & 16.62  & 17.00 & 15.59 &  0.0025          \\ 
 4.5 &  16.48 & 16.90 & 16.98&  16.66  & 16.87 & 16.38  &   0.001       \\ 
 5   &  16.58 & 16.73 & 17.04& 17.22   & 17.03 & 16.91  &  0.0005        \\ 
 5.5 & 16.66 & 16.92 &  \textbf{17.25}&  \textbf{17.33}  & 16.91 & 16.96 & 0.00025 \\ 
 6   & 16.73 & 17.24 & 17.23& 17.23  & 16.86 & \textbf{17.10} &  0.0001\\ 
 6.5 & 16.82 & 17.30 & 17.16& 17.17  & 16.96  & 17.07 & 0.00005 \\ 
 7   & \textbf{16.87}  & \textbf{17.32}  & 17.20& 17.23  & 16.70  & 16.70 & 0.00001\\ 
\bottomrule[1pt]
\end{tabular}
\caption{The S/N values on chunk 2 of mat 0 for different thresholds and different RFI removal methods. The S/N of the raw data is 16.92. } 
\label{table:comparison_chunk_2}
\end{table}

\begin{table} 
\centering 
\begin{tabular}{ cccccccc } 
\toprule[1pt] 
 \multicolumn{8}{c}{ chunk 3 of mat 0}  \\
  \midrule[0.8pt] 
 Th  & MAD   & SE   & SRE$_{s}$ & SRE$_{a}$ & SK    & SW    & $\alpha$-level  \\ 
 \midrule[0.8pt]   \midrule[0.8pt] 
 3   & 13.26 & 16.18 & 15.96& 16.39   & 16.00 & 15.17 & 0.01            \\ 
 3.5 & 14.60 & 16.26 &  15.76& 15.62   & 16.11 & 15.47 & 0.005        \\ 
 4   & 15.29 & 16.21 &  16.03& 15.97   & 16.06 & 15.79 & 0.0025       \\ 
 4.5 & 15.56 & 16.36 & 15.56 &  16.63  & 16.52 & 15.88  &   0.001       \\ 
 5   & 15.82 & 16.41 & 16.70& 16.85   & \textbf{16.58} & 16.25    &   0.0005        \\ 
 5.5 & 15.99 & \textbf{16.43}  & \textbf{16.92} & 16.85   & 16.55 & 16.27 & 0.00025   \\ 
 6   & \textbf{16.04} & 16.01 & 16.78& \textbf{16.87}   & 16.56 & 16.36 & 0.0001 \\ 
 6.5 & 16.04 & 15.81 & 16.60& 16.69   & 16.56 & \textbf{16.38} & 0.00005 \\ 
 7   & 16.03 & 15.68 & 16.75& 16.67   & 16.43 &16.33 & 0.00001 \\ 
\bottomrule[1pt]
\end{tabular}
\caption{The S/N values on chunk 3 of mat 0 for different thresholds and different RFI removal methods. The S/N of the raw data is 16.03. } 
\label{table:comparison_chunk_3}
\end{table}

\begin{table} 
\centering 
\begin{tabular}{ cccccccc } 
\toprule[1pt]
 \multicolumn{8}{c}{ chunk 4 of mat 0 } \\
  \midrule[0.8pt] 
 Th  & MAD   & SE   & SRE$_{s}$ & SRE$_{a}$ & SK    & SW    & $\alpha$-level  \\ 
\midrule[0.8pt]  \midrule[0.8pt] 
 3   & 13.66 & 16.45 & 15.42& 15.41   & 16.25 & 16.12 & 0.01            \\ 
 3.5 & 15.51 & \textbf{16.96} & 16.03& 16.50  & 16.80 & 16.70 &  0.005          \\ 
 4   & 16.32 & 16.61 & 16.72&  16.49  & \textbf{17.25} & \textbf{16.81} &   0.0025        \\ 
 4.5 & 16.65 & 16.33 & 16.51& 16.57  & 17.10 & 16.52 &  0.001       \\ 
 5   & 16.78 & 16.43 & 16.63&  16.58  & 16.84 & 16.43 &   0.0005        \\ 
 5.5 & 16.94 & 16.42 & 16.61& 16.57   & 16.64 & 16.57 & 0.00025   \\ 
 6   & 16.94 & 16.58 & 16.76& 16.65   & 16.62 & 16.43 &  0.0001  \\ 
 6.5 & \textbf{16.95} & 16.22 & \textbf{16.81}&  16.70  & 16.37 & 16.48 &  0.00005 \\ 
 7   & 16.95 & 16.26 & 16.74 & \textbf{16.70}  & 16.55 & 16.75 & 0.00001 \\ 
\bottomrule[1pt]
\end{tabular}
\caption{The S/N values on chunk 4 of mat 0 for different thresholds and various RFI removal methods. The S/N of the raw data is 16.93. } 
\label{table:comparison_chunk_4}
\end{table}

\subsection{Performance Analysis of $mat \ 2$}
\begin{figure}
\begin{center}    
\begin{tabular}{c}
\includegraphics[width=0.45\textwidth]{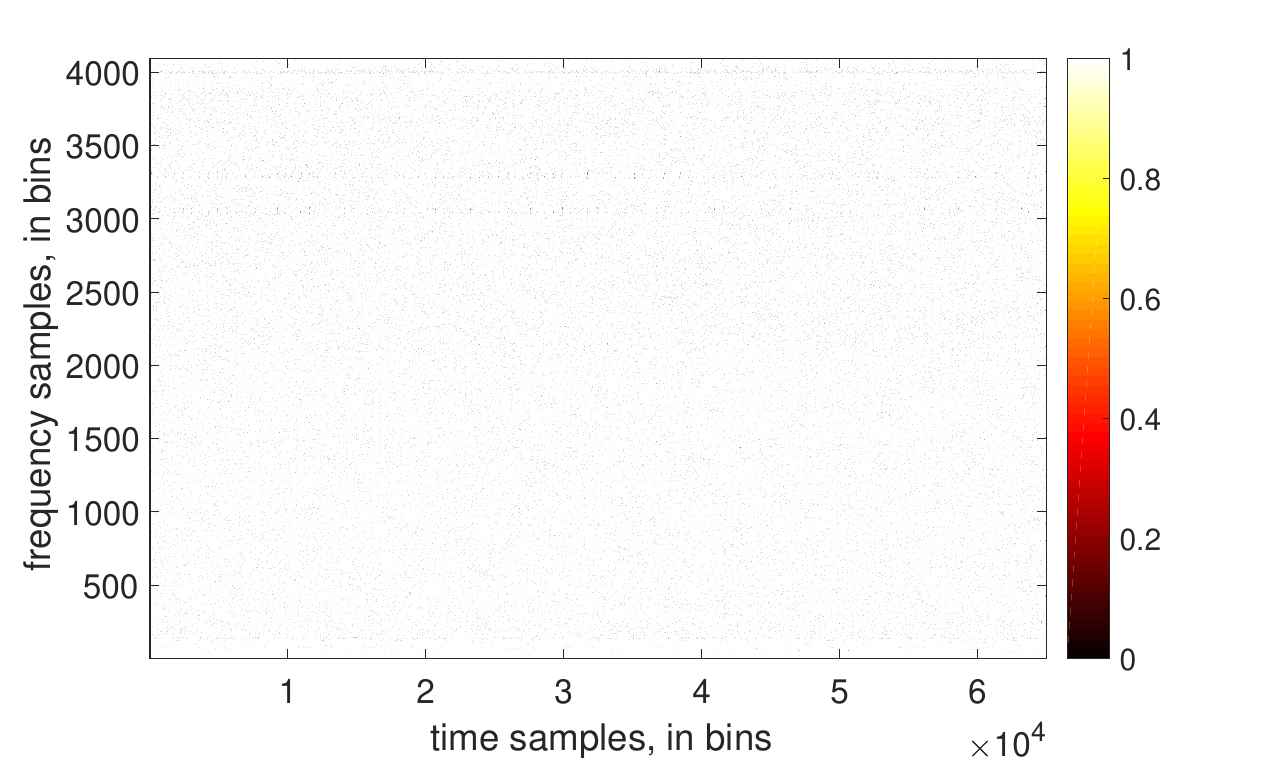} 
\end{tabular}
\caption{The mask generated by MAD at the threshold of $3\sigma$ when the MAD method is applied to $chunk \ 2$ of $mat \  2$. For further details, see Fig.~7.}
\label{fig:MAD_mask_mat_2_chunk_2_Th_3}
\end{center}
\end{figure}

\begin{figure}
\begin{center}    
\begin{tabular}{c}
\includegraphics[width=0.45\textwidth]{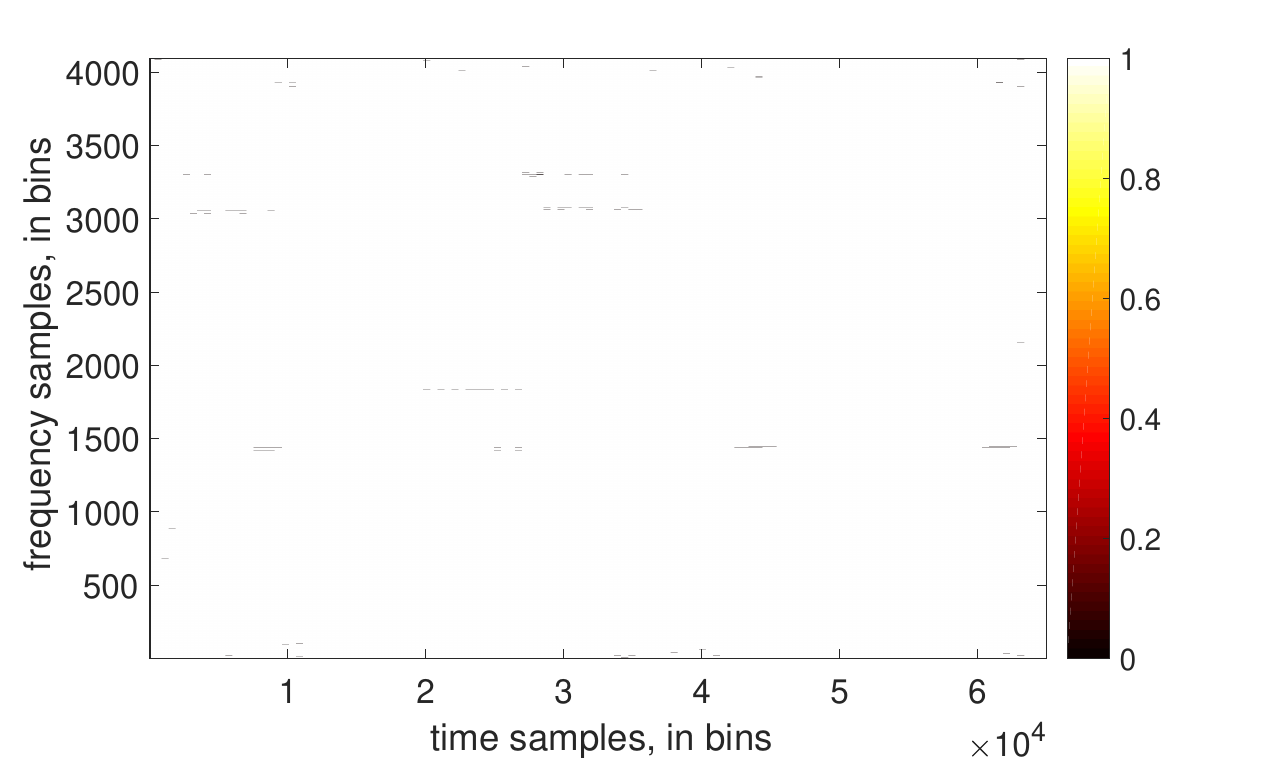} 
\end{tabular}
\caption{The mask generated by SE at the threshold of $6\sigma$ when the SE method is applied to $chunk \ 2$ of $mat \ 2$. For further details, see Fig.~7.}
\label{fig:SE_mask_mat_2_chunk_2_Th_6}
\end{center}
\end{figure}

\begin{figure}
\begin{center}    
\begin{tabular}{c}
\includegraphics[width=0.45\textwidth]{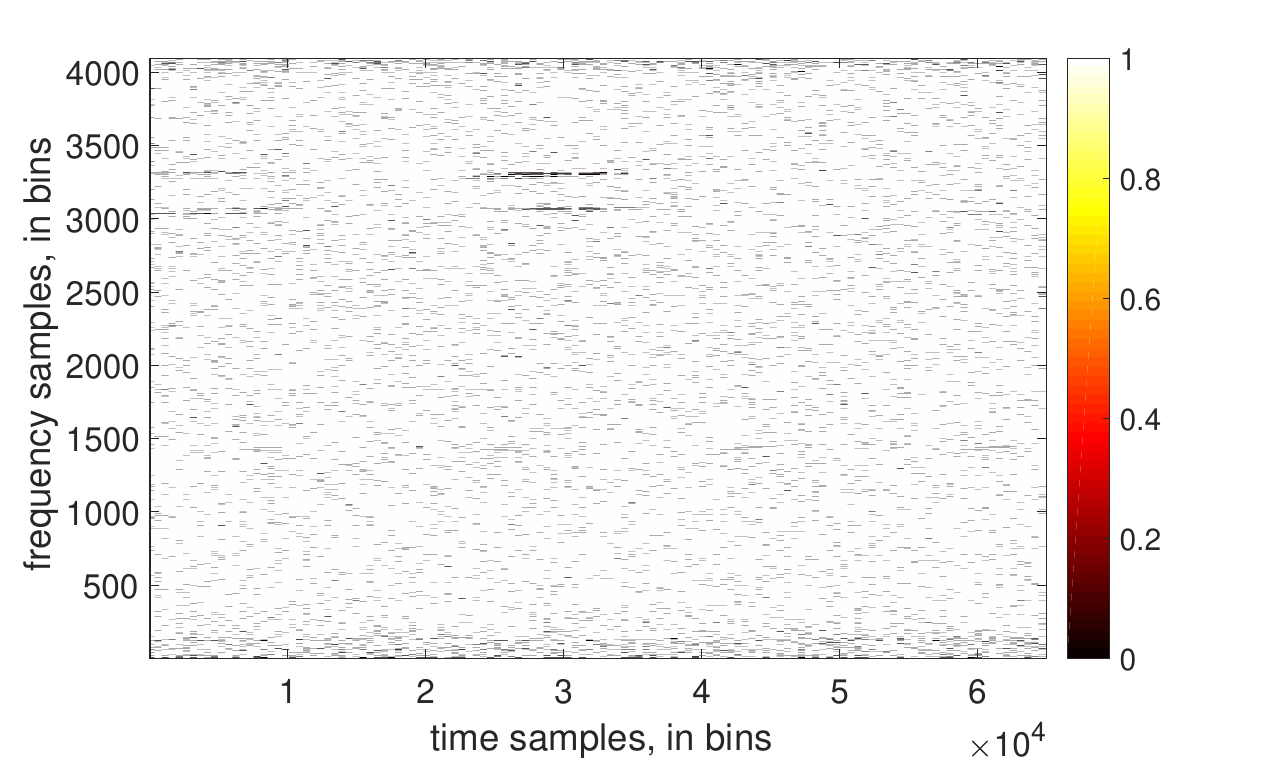} 
\end{tabular}
\caption{The mask generated by symmetrical SRE at the threshold of $3\sigma$ when the SRE method is applied to $chunk \ 2$ of $mat \ 2$. For further details, see Fig.~7.}
\label{fig:SRE_s_mask_mat_2_chunk_2_Th_3}
\end{center}
\end{figure}

\begin{figure}
\begin{center}    
\begin{tabular}{c}
\includegraphics[width=0.45\textwidth]{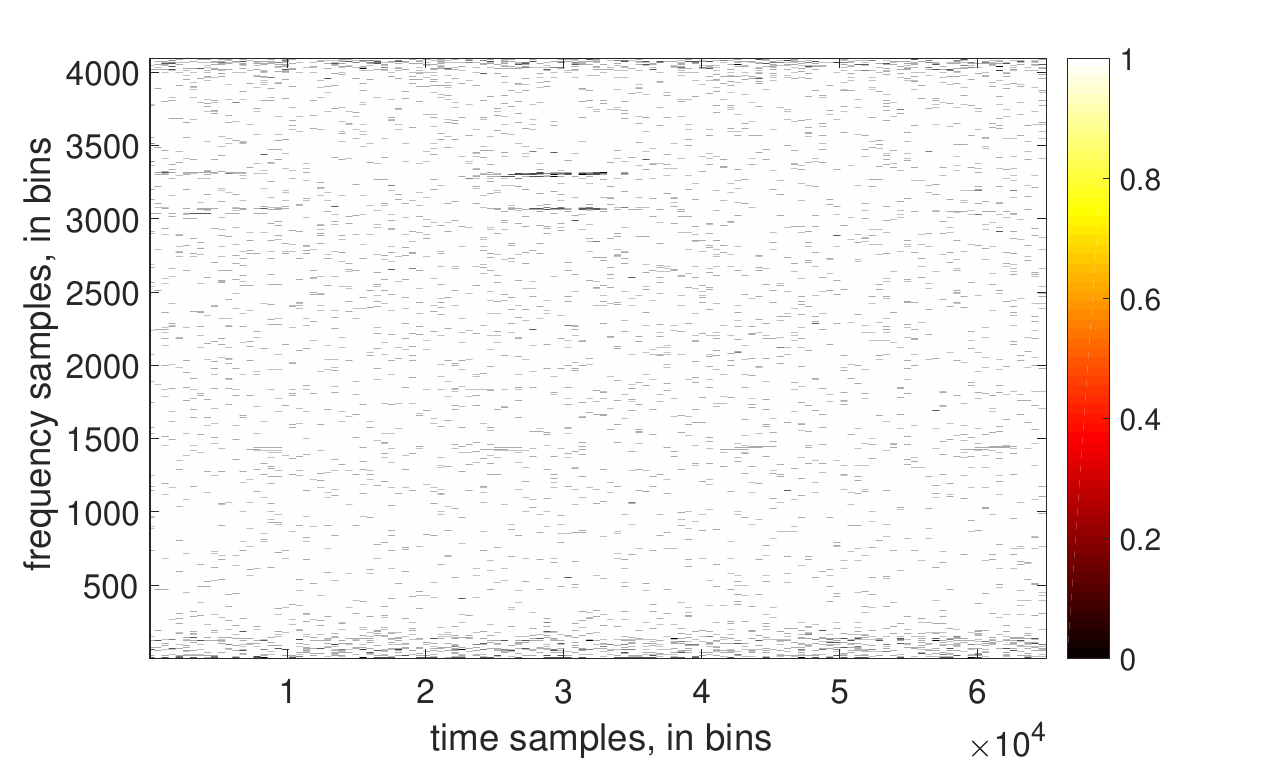} 
\end{tabular}
\caption{The mask generated by asymmetrical SRE at the threshold of $3.5\sigma$ when the asymmetrical SRE method is applied to $chunk \ 2$ of $mat \ 2$. For further details, see Fig.~7.}
\label{fig:SRE_a_mask_mat_2_chunk_2_Th_3_5}
\end{center}
\end{figure}

\begin{figure}
\begin{center}    
\begin{tabular}{c}
\includegraphics[width=0.45\textwidth]{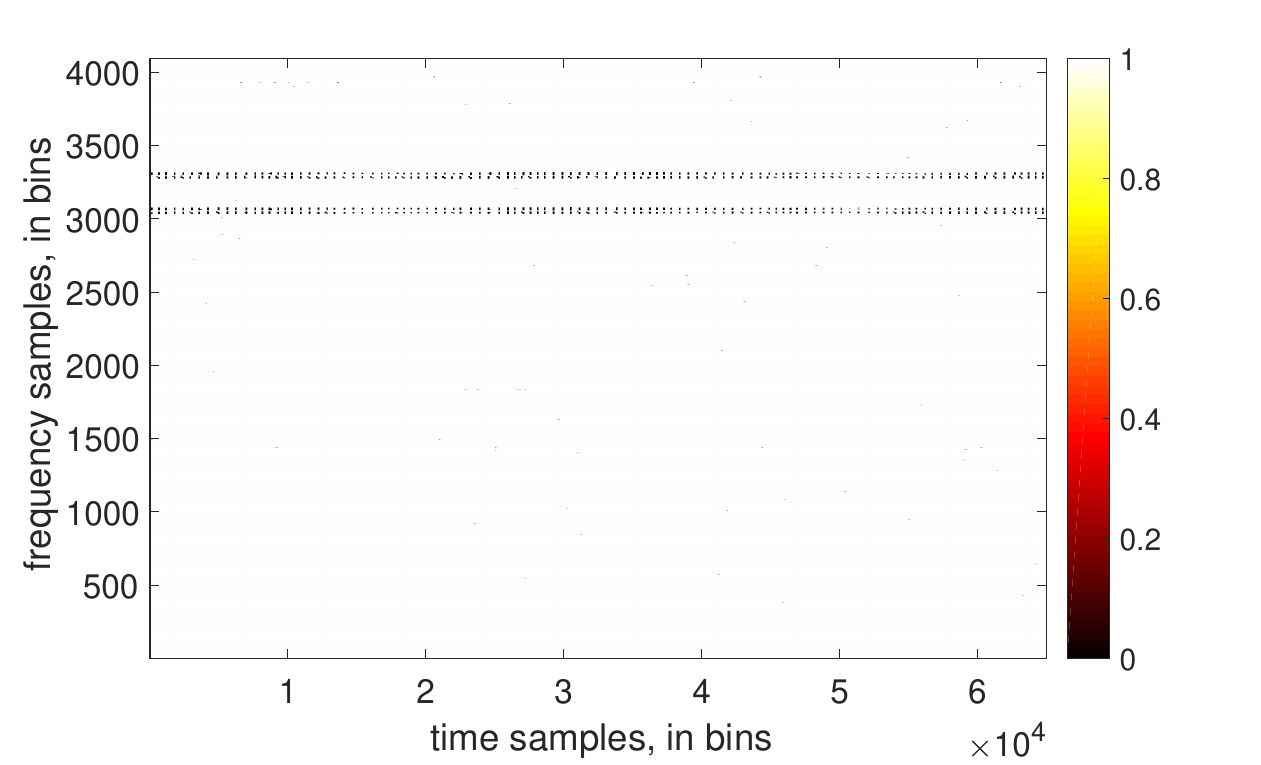} 
\end{tabular}
\caption{The mask generated by SK at the value of threshold set to $7\sigma$ when the SK method is applied to $chunk \ 2$ of $mat \ 2$. Note how well SK detects the RFI signals of varying frequency in the frequency range between 3000 and 3500. For further details, see Fig.~7.}
\label{fig:SK_mask_mat_2_chunk_2_Th_7}
\end{center}
\end{figure}

\begin{figure}
\begin{center}    
\begin{tabular}{c}
\includegraphics[width=0.45\textwidth]{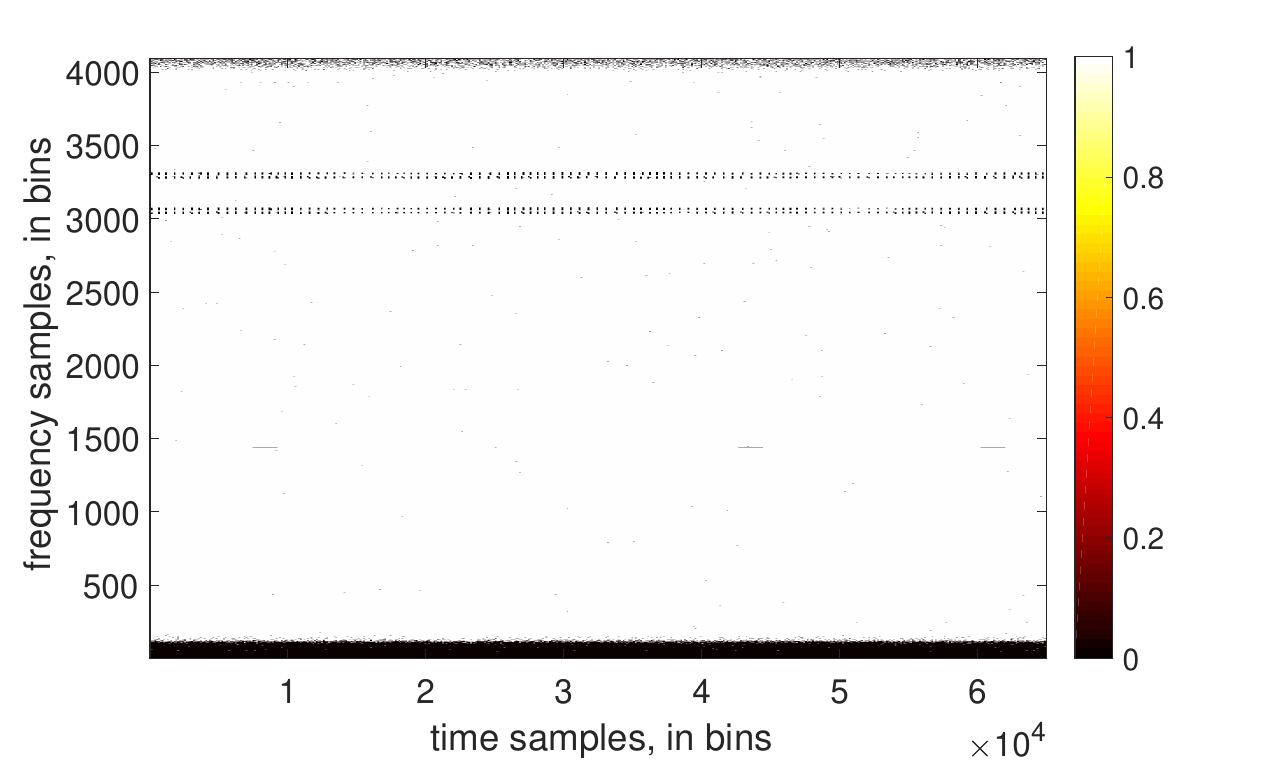} 
\end{tabular}
\caption{The mask generated by SW at the $\alpha$ level of 0.0001 when the SW method is applied to $chunk \ 2$ of $mat \ 2$. Similar to SK, SW detects RFI signals of varying frequency in frequency channels between 3000 and 3500. For further details, see Fig.~7.}
\label{fig:SW_mask_mat_2_chunk_2_alpha_0_0001}
\end{center}
\end{figure}

\begin{figure}
\centering
\includegraphics[width=0.45\textwidth]{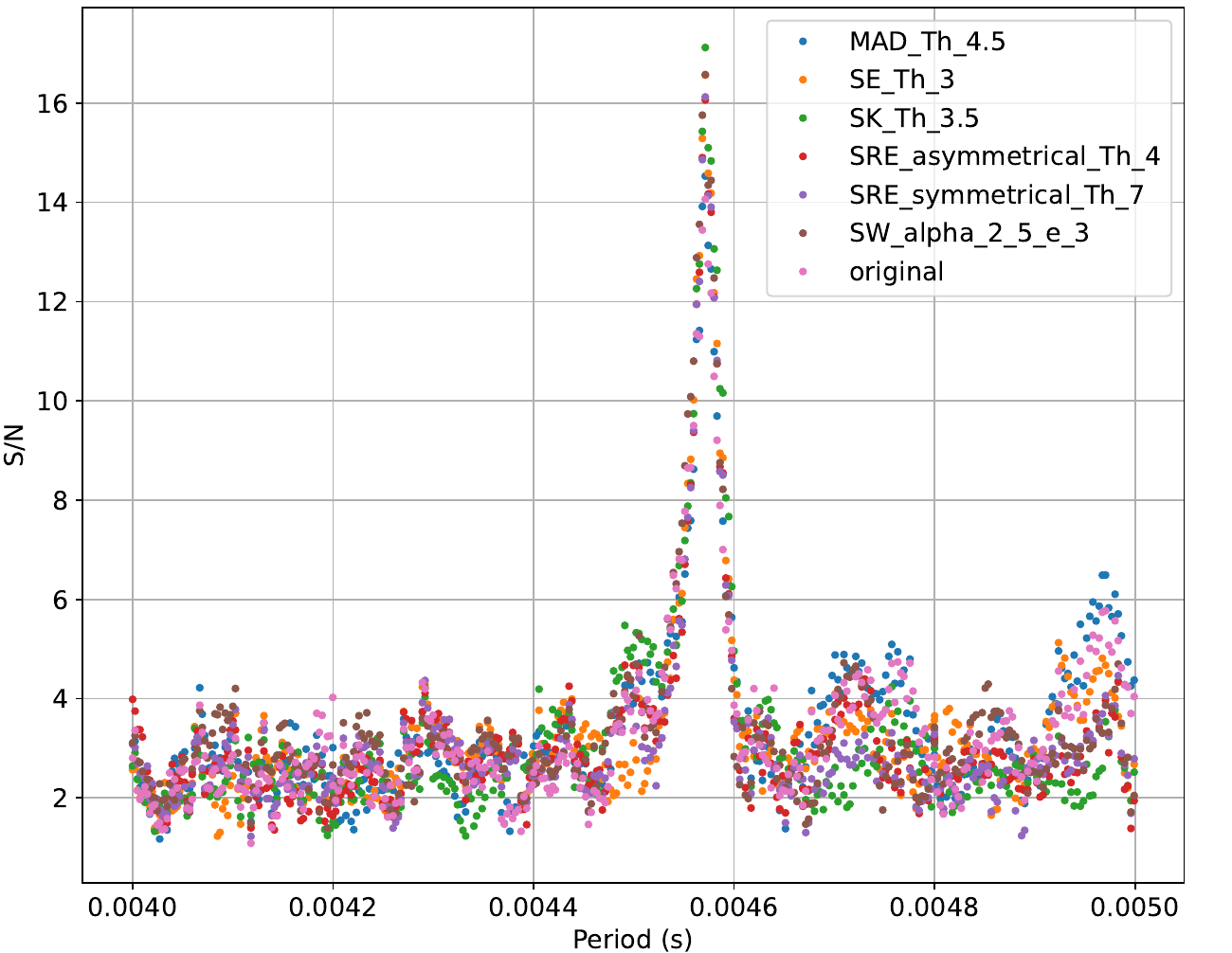} 
\caption{Periodogram results from {\sc riptide} for and baseline RFI detection methods obtained from the data in $chunk \ 0$ of $mat \ 2.$}
\label{fig:chun0_mat2_SN}
\end{figure}

\begin{figure}
\centering
\includegraphics[width=0.45\textwidth]{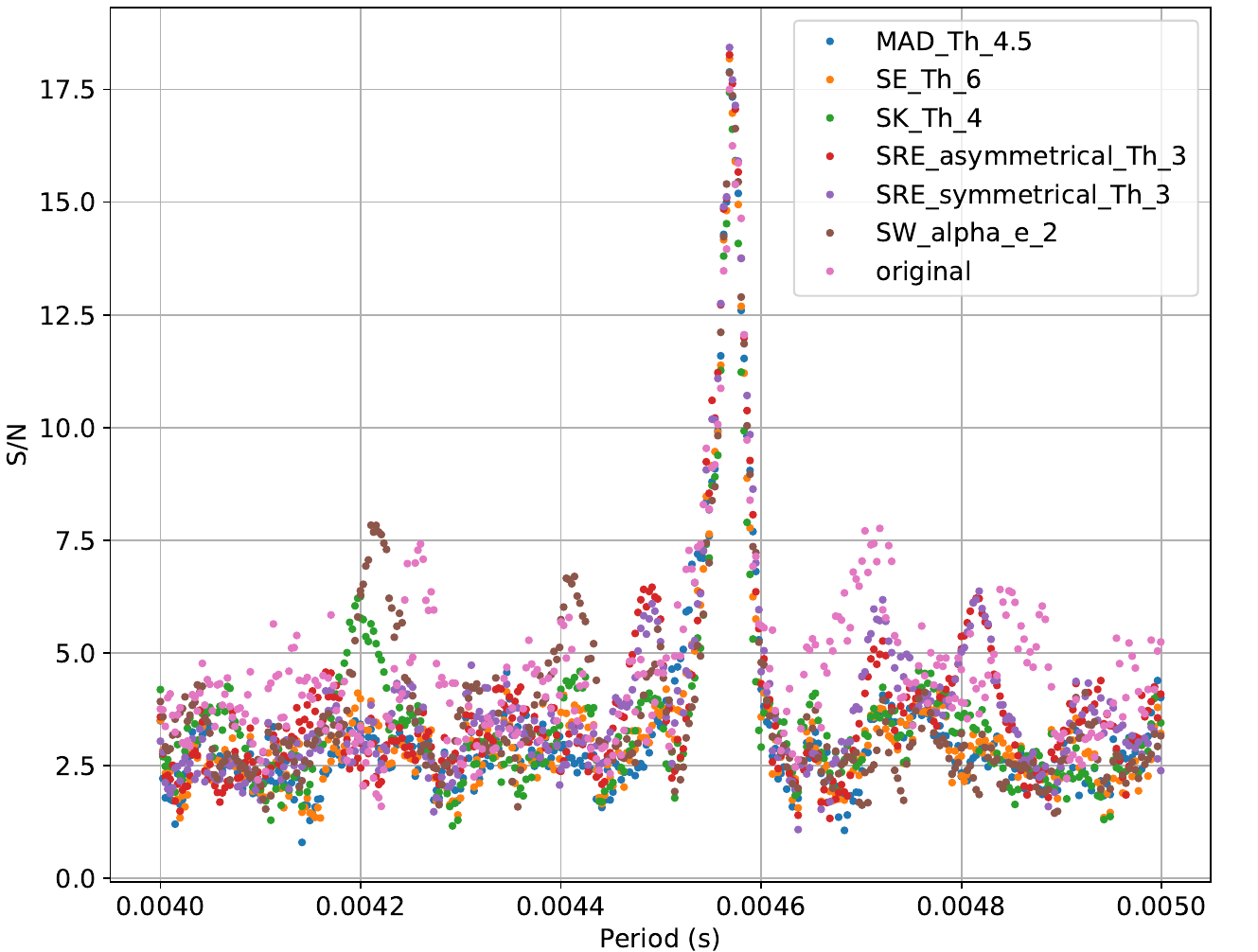} 
\caption{Periodogram results from {\sc riptide} for and baseline RFI detection methods obtained from the data in $chunk \ 1$ of $mat \ 2.$}
\label{fig:chun1_mat2_SN}
\end{figure}

\begin{figure}
\centering
\includegraphics[width=0.45\textwidth]{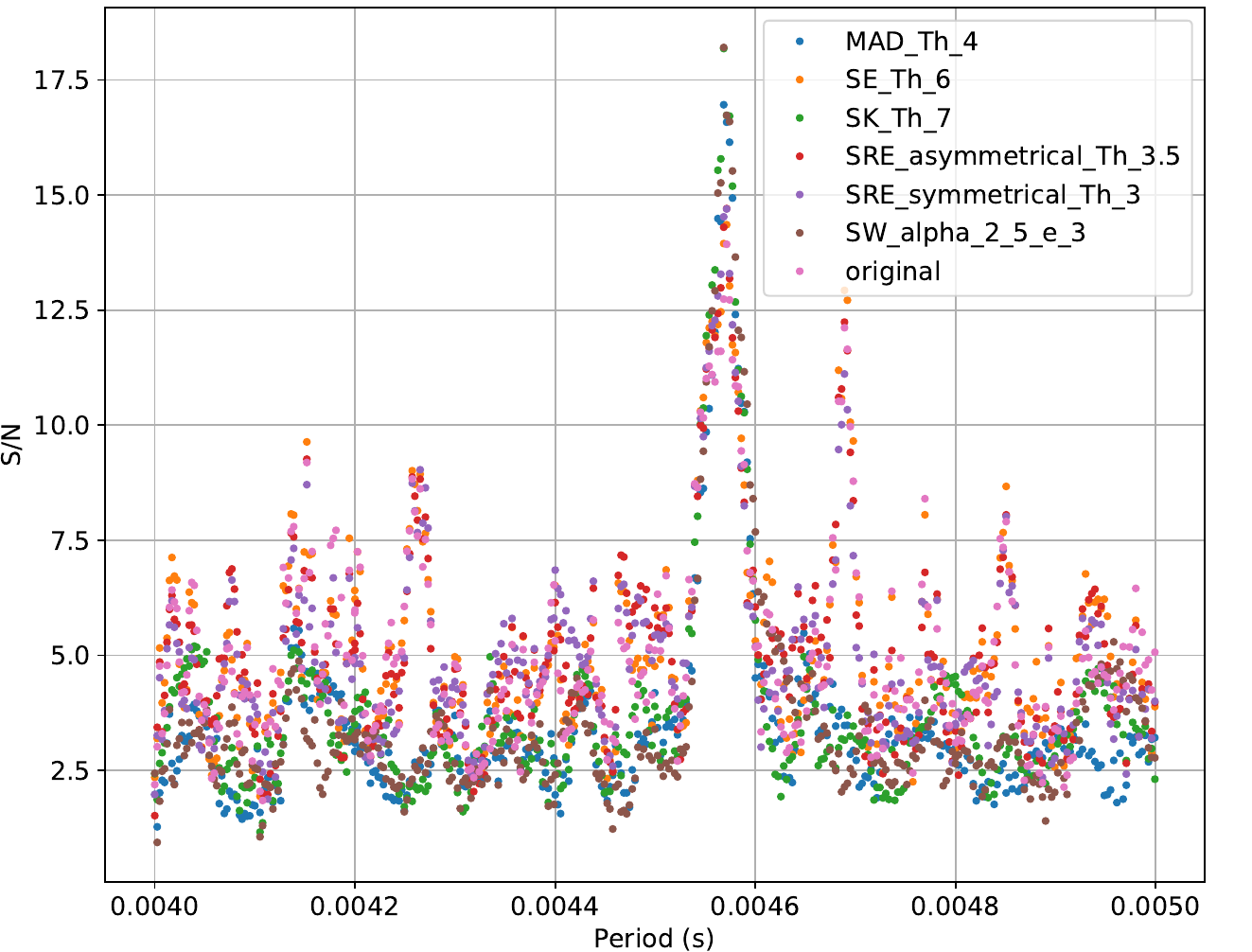} 
\caption{Periodogram results from {\sc riptide} for and baseline RFI detection methods obtained from the data in $chunk \ 2$ of $mat \ 2.$}
\label{fig:chun2_mat2_SN}
\end{figure}

\begin{figure}
\centering
\includegraphics[width=0.45\textwidth]{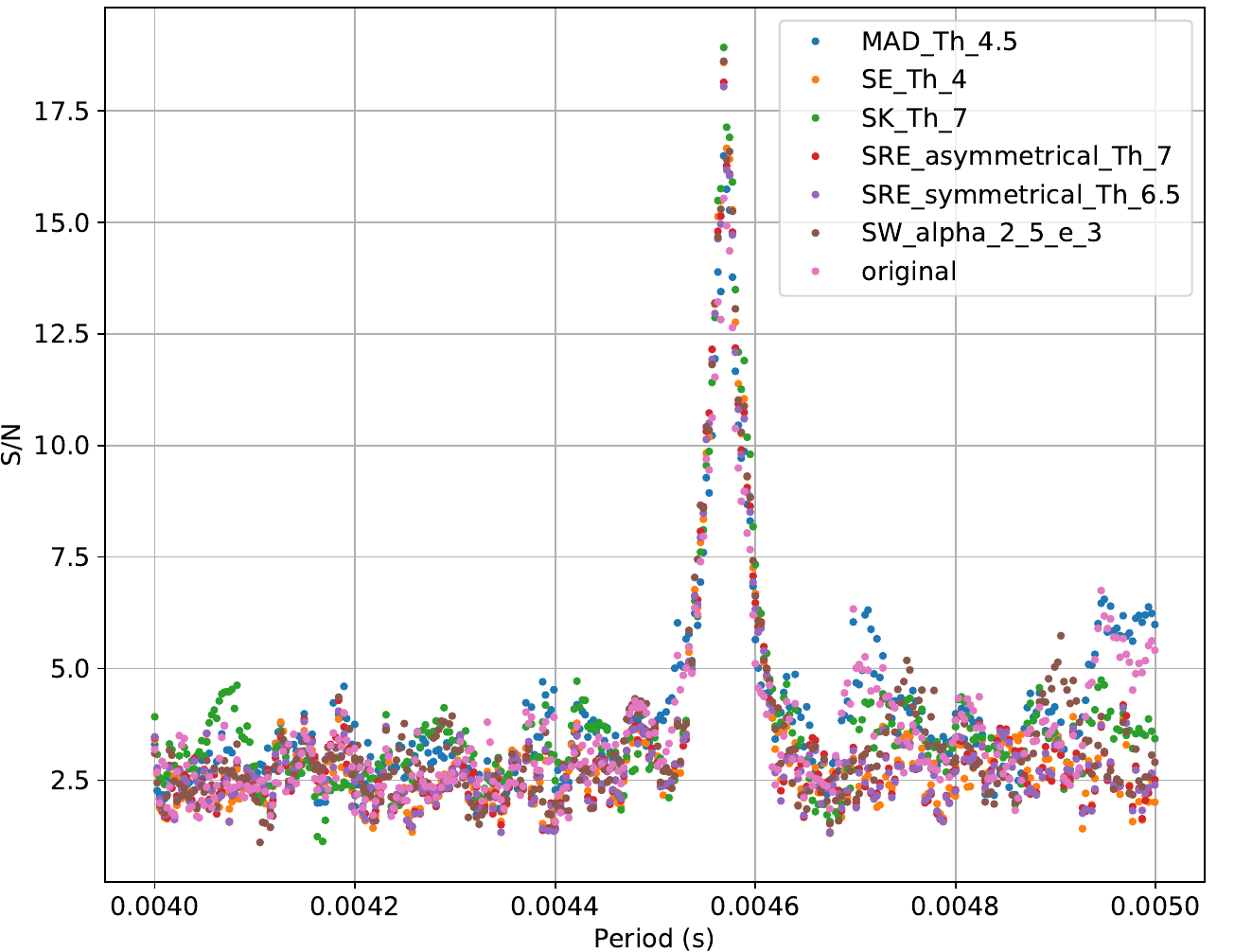} 
\caption{Periodogram results from {\sc riptide} for and baseline RFI detection methods obtained from the data in $chunk \ 3$ of $mat \ 2.$}
\label{fig:chun3_mat2_SN}
\end{figure}

\begin{figure}
\centering
\includegraphics[width=0.45\textwidth]{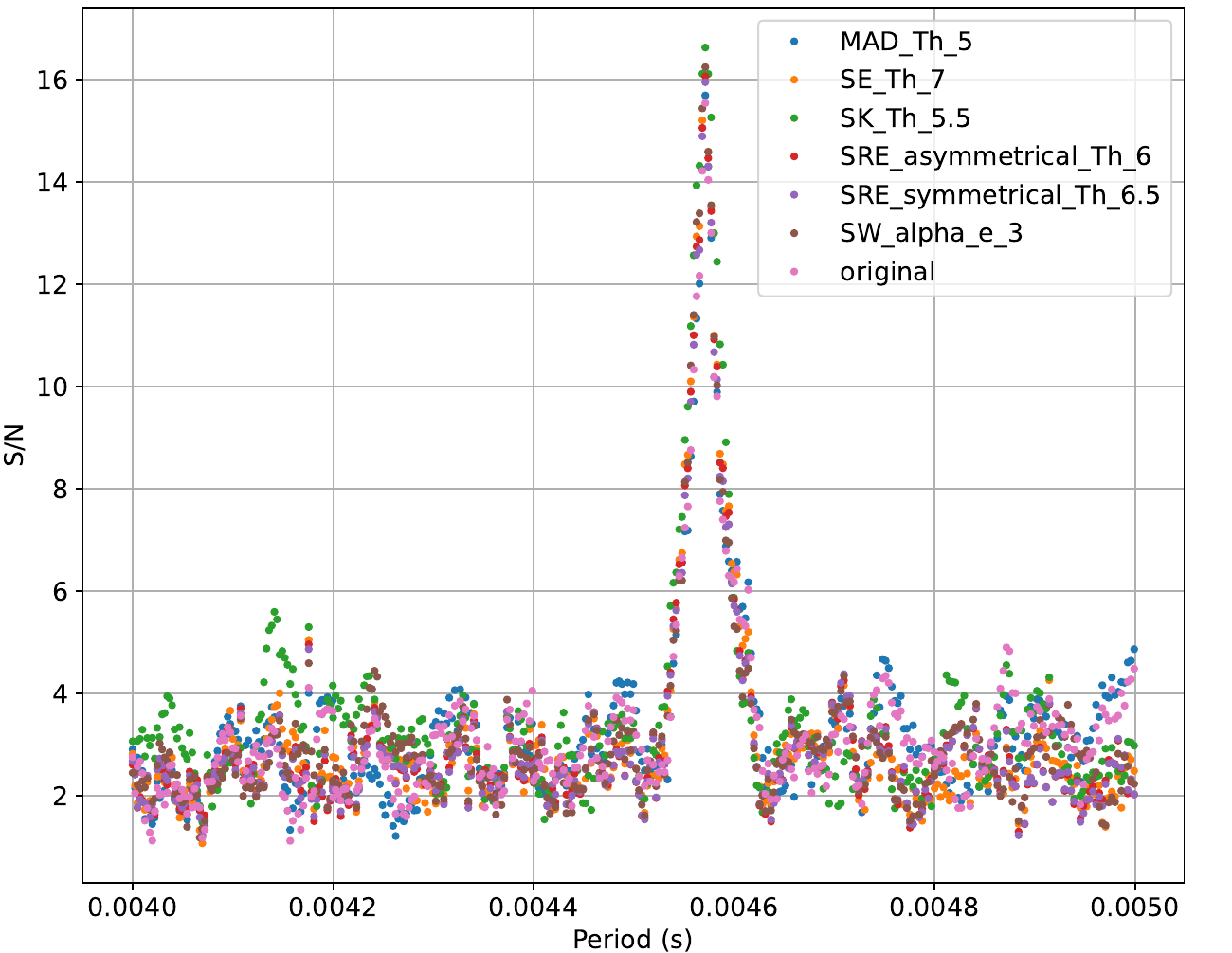} 
\caption{Periodogram results from {\sc riptide} for and baseline RFI detection methods obtained from the data in $chunk \ 4$ of $mat \ 2.$}
\label{fig:chun4_mat2_SN}
\end{figure}

The data in $mat \ 2$ contain another challenging type of RFI, a strong signal varying in frequency and time. Figures \ref{fig:SK_mask_mat_2_chunk_2_Th_7} and  \ref{fig:SW_mask_mat_2_chunk_2_alpha_0_0001} demonstrate the ability of SK and SW methods to detect and flag this type of RFI in $chunk \ 2$ of $mat \ 2.$ The other RFI detection methods miss to flag this type of RFI signal which is shown in Figures \ref{fig:MAD_mask_mat_2_chunk_2_Th_3}, \ref{fig:SE_mask_mat_2_chunk_2_Th_6}, \ref{fig:SRE_s_mask_mat_2_chunk_2_Th_3}, and \ref{fig:SRE_a_mask_mat_2_chunk_2_Th_3_5}.  The S/N values for all chunks of $mat \ 2$ are displayed in Tables \ref{table:comparison_mat_2_chunk_0} through \ref{table:comparison_mat_2_chunk_4}. Note that for $chunk \ 2$ flagging the RFI signal varying in frequency and time resulted in considerably improved S/N values for SK and SW compared to the S/N value of raw data. The plots of a single folded pulse for the choice of the best S/N value for the six RFI detection methods as well as for the case of raw data are provided in Figures \ref{fig:chun0_mat2_SN} through \ref{fig:chun4_mat2_SN} for $chunk \ 0$ through $4$ of $mat \ 2,$ respectively.      

\begin{table}
\centering 
\begin{tabular}{ cccccccc } 
\toprule[1pt] 
 \multicolumn{7}{c}{ chunk 0 of mat 2 } \\
 \midrule[0.8pt]  
 Th  & MAD   & SE   & SRE$_{s}$  & SRE$_{a}$ & SK    & SW    & $\alpha$-level  \\ 
 \midrule[0.8pt]   \midrule[0.8pt]  
 3   & 12.64 & \textbf{16.56}  & 15.72& 15.56  & 16.54 & 14.31 & 0.01            \\ 
 3.5 & 14.10 & 16.41 & 15.86 & 15.89 & \textbf{17.12}  & 15.39 & 0.005           \\ 
 4   & 14.43 &  16.25  & 16.06 & \textbf{16.06} & 16.48 & \textbf{16.57} & 0.0025           \\ 
 4.5 & \textbf{14.53} & 16.11 & 15.95& 15.99   & 16.48 & 16.32 & 0.001         \\ 
 5   & 14.52 & 16.36 & 15.98 & 15.84 & 16.52 & 16.03 & 0.0005          \\ 
 5.5 &  14.42  & 16.38 & 16.11&  16.01  & 16.34 & 16.40 & 0.00025 \\ 
 6   & 14.31 & 16.46 & 16.07& 16.05 & 15.78 &16.10 & 0.0001  \\ 
 6.5 & 14.18 & 16.40 & 15.99& 16.06  & 15.86 &16.08 &0.00005  \\ 
 7   & 14.12 & 15.58 & \textbf{16.12}& 16.05  & 15.78 &16.17 &0.00001 \\ 
\bottomrule[1pt]
\end{tabular}
\caption{The S/N values on $chunk \ 0$ of $mat \ 2$ for different thresholds and different RFI removal methods. The S/N of the raw data is 14.06. } 
\label{table:comparison_mat_2_chunk_0}
\end{table}
\begin{table}
\centering 
\begin{tabular}{ cccccccc } 
\toprule[1pt] 
 \multicolumn{7}{c}{ chunk 1 of mat 2} \\
\midrule[0.8pt]
  Th  & MAD   & SE   & SRE$_{s}$  & SRE$_{a}$ & SK    & SW    & $\alpha$-level  \\ 
\midrule[0.8pt] \midrule[0.8pt]
 3   & 15.68 & 16.36 & \textbf{18.43}& \textbf{18.27}  & 17.00 & \textbf{17.87} & 0.01            \\ 
 3.5 & 17.17 & 17.47 & 17.93& 17.60  & 17.21 & 16.97 & 0.005           \\ 
 4   & 17.70 & 17.92 & 17.68& 17.90  & \textbf{17.44} & 16.51 & 0.0025           \\ 
 4.5 & \textbf{17.89} & 17.89 & 17.95&  17.89 & 17.27 & 17.13 & 0.001          \\ 
 5   & 17.83 & 17.94 & 17.75&  17.81 & 17.37 &17.41  & 0.0005          \\ 
 5.5 & 17.77 & 18.01 & 18.01& 17.94  & 17.07 &17.44 & 0.00025        \\ 
 6   & 17.75 & \textbf{18.18} & 17.97& 17.83  & 16.99 &17.51 &0.0001           \\ 
 6.5 & 17.75 & 18.00 & 17.88& 17.96  & 17.41 &17.61 &0.00005          \\ 
 7   & 17.73 & 18.05 & 17.88& 17.90  & 17.06 &17.70 &0.00001          \\ 
\bottomrule[1pt]
\end{tabular}
\caption{The S/N values on $chunk \ 1$ of $mat \ 2$ for different thresholds and different RFI removal methods. The S/N of the raw data is 17.50. } 
\label{table:comparison_mat_2_chunk_1}
\end{table}
\begin{table}
\centering 
\begin{tabular}{ cccccccc } 
\toprule[1pt]  
 \multicolumn{7}{c}{ chunk 2 of mat 2} \\
\midrule[0.8pt]  
 Th  & MAD   & SE   & SRE$_{s}$  & SRE$_{a}$ & SK    & SW    & $\alpha$-level  \\ 
\midrule[0.8pt]  \midrule[0.8pt]  
  3   & 15.21 & 13.46 & \textbf{14.70}& 14.23  & 17.92 & 18.17 & 0.01            \\ 
 3.5 & 16.61 & 13.63 & 14.25& \textbf{14.70}  & 17.71 & 17.97 & 0.005           \\ 
 4   & \textbf{16.96} & 13.83 & 14.09& 13.60  & 17.51 & \textbf{18.20}  & 0.0025          \\ 
 4.5 & 16.88 & 13.95 & 14.08& 13.95  & 17.73 & 17.89 & 0.001          \\ 
 5   & 16.84 & 14.12 & 14.09& 13.99  & 17.43 & 17.67 & 0.0005          \\ 
 5.5 & 16.68 & 14.34 & 13.98& 14.05  & 17.72 & 17.71 & 0.00025    \\ 
 6   & 16.48 & \textbf{14.36} & 13.87 &  13.87 & 17.72 &17.71 &0.0001  \\ 
 6.5 & 16.31 & 14.21 & 13.81& 13.88  & 17.66 &17.71 &0.00005   \\ 
 7   & 16.16 & 14.06 & 13.86& 13.87  & \textbf{18.18} &17.57 &0.00001 \\ 
\bottomrule[1pt]
\end{tabular}
\caption{The S/N values on $chunk \ 2$ of $mat \ 2$ for different thresholds and different RFI removal methods. The S/N of the raw data is 13.93. } 
\label{table:comparison_mat_2_chunk_2}
\end{table}
\begin{table}
\centering 
\begin{tabular}{ cccccccc } 
\toprule[1pt] 
 \multicolumn{7}{c}{ chunk 3 of mat 2} \\
\midrule[0.8pt]  
 Th  & MAD   & SE   & SRE$_{s}$  & SRE$_{a}$ & SK    & SW    & $\alpha$-level  \\ 
\midrule[0.8pt]   \midrule[0.8pt]  
 3   & 14.36 & 18.43 & 16.45& 16.61  & 18.42 & 18.25 & 0.01            \\ 
 3.5 & 15.79 & 18.49 & 16.64&  16.94  & 17.92 & 18.33 & 0.005           \\ 
 4   & 16.46 & \textbf{18.58} & 17.24& 17.09  & 18.44 & 18.34 & 0.001           \\ 
 4.5 & \textbf{16.48} & 18.39 & 17.45& 17.42  & 18.81 & \textbf{18.60} &  0.0025          \\ 
 5   & 16.48 & 18.27 & 17.66&  17.82 & 18.07 & 18.27 & 0.0005          \\ 
 5.5 & 16.16& 18.14 & 17.84&  17.97 & 18.24 &18.28 & 0.00025 \\ 
 6   & 16.09 & 18.12 & 17.97& 18.07  & 18.90 &18.15 & 0.0001 \\ 
 6.5 & 16.07 & 18.14 & \textbf{18.04}& 18.04  & 18.88 &18.15 & 0.00005\\ 
 7   & 16.01 & 18.11 & 18.03& \textbf{18.13}  & \textbf{18.91} &18.18 & 0.00001\\ 
 \bottomrule[1pt]
\end{tabular}
\caption{The S/N values on Chunk 3 of mat 2 for different thresholds and different RFI removal methods. The S/N of the raw data is 15.53. } 
\label{table:comparison_mat_2_chunk_3}
\end{table}
\begin{table}
\centering 
\begin{tabular}{ cccccccc } 
\toprule[1pt]  
 \multicolumn{7}{c}{ chunk 4 of mat 2} \\
 \midrule[0.8pt]  
 Th  & MAD   & SE   & SRE$_{s}$  & SRE$_{a}$ & SK    & SW    & $\alpha$-level  \\ 
 \midrule[0.8pt]    \midrule[0.8pt]  
  3   & 13.64 & 15.55 & 14.81& 14.52  & 15.76 & 14.13 & 0.01            \\ 
 3.5 & 15.07 & 15.61 & 15.38&  14.93 & 16.22 & 15.64 &  0.005           \\ 
 4   & 15.49 & 15.83 & 15.38& 15.73  & 16.37 & 16.05 & 0.0025           \\ 
 4.5 & 15.65 & 15.89 & 15.70& 15.81  & 16.49 & \textbf{16.25} & 0.001       \\ 
 5   & \textbf{15.69} & 15.91 & 15.77& 15.85  & 16.61 & 15.99 & 0.0005        \\ 
 5.5 & 15.62 & 15.87 & 15.92&  15.85 & \textbf{16.63} &16.04 & 0.00025    \\ 
 6   & 15.62 & 16.03 & 15.91&  \textbf{16.06} & 16.53 &16.00 &0.0001     \\ 
 6.5 & 15.62 & 16.00 & \textbf{15.95}& 15.98  & 16.50 &16.14 & 0.00005   \\ 
 7   & 15.59 & \textbf{16.10} & 15.87&  15.94 & 16.46 &16.06 & 0.00001    \\ 
\bottomrule[1pt]
\end{tabular}
\caption{The S/N values on $chunk \ 4$ of $mat \ 2$ for different thresholds and various RFI removal methods. The S/N of the raw data is 15.54. } 
\label{table:comparison_mat_2_chunk_4}
\end{table}


%
%
%
%

\subsection{General observations}
To summarize the performance of the tested methods for the detection and flagging RFI signals in astronomy data, the following general observations are made.  
\begin{enumerate}
\item In every analyzed case, the application of SE, symmetrical SRE, asymmetrical SRE, SK, and SW resulted in an improved value of S/N compared to the S/N of the raw data. Unlike the five methods above, the application of MAD on many occasions leads to a reduced value of S/N compared to the S/N of the raw data. 
\item SE, symmetric SRE, asymmetric SRE, SK, and SW showcase their ability to detect broadband RFI signals (e.g., $chunk \ 0$ in $mat \ 0$). 
\item Varying in frequency and time RFI signals are best detected by SK and SW tests (see RFI in $chunk \ 2$ of $mat \ 2$) as well. 
\item Raw channelized voltages yielding a high S/N of the folded pulse do not benefit from RFI detection and flagging methods. 
\item Asymmetric SRE perfoms better than symmetric SRE. 
\end{enumerate} 

\section{Conclusions}
\label{sec:summary}
The range of statistical methods examined in this work is used as an indicator of how clean, RFI-free Gaussian distributed complex-valued frequency channel characteristics vary from the characteristics of RFI-contaminated channels. A demonstration of typical RFI environments was explored by applying Median Absolute Deviation (MAD), Spectral Entropy (SE), Spectral Relative Entropy (SRE), Spectral Kurtosis (SK), and Shapiro-Wilks (SW) test for Normality to complex-valued channelized voltage data collected with the GBT.

The S/N of a single folded pulse was selected as a means to compare the performance of the RFI detection methods. The application of MAD, SE, SRE, SK, and SW on the millisecond pulsar data of J1713+0747 illustrates that MAD does not always filter RFI effectively. Both MAD, SE, and SRE often keep the same RFI artifacts that are found in the original data. SK and SW successfully detect and remove both broadband RFI signals and signals varying in frequency. All of the RFI detection tests except MAD increase the S/N of the pulsar data. In the future, further investigations of these methods on larger data sets are strongly encouraged.

\section*{Acknowledgements}
{This research is partially supported by the National Science Foundation under Awards No. AST-2307581, the Natural Science Foundation of China (NSFC No. 61906149), and the Natural Science Foundation of Chongqing (cstc2021jcyj-msxmX1068). The authors would also like to thank their colleagues at the Green Bank Observatory and West Virginia University for providing the data set used throughout this research. }

\section*{Data Availability}
The data used in this study are available upon request.  
 

\bibliographystyle{mnras}
\bibliography{frb_refs}

\begin{thebibliography}{}
\makeatletter
\relax
\def\mn@urlcharsother{\let\do\@makeother \do\$\do\&\do\#\do\^\do\_\do\%\do\~}
\def\mn@doi{\begingroup\mn@urlcharsother \@ifnextchar [ {\mn@doi@}
  {\mn@doi@[]}}
\def\mn@doi@[#1]#2{\def\@tempa{#1}\ifx\@tempa\@empty \href
  {http://dx.doi.org/#2} {doi:#2}\else \href {http://dx.doi.org/#2} {#1}\fi
  \endgroup}
\def\mn@eprint#1#2{\mn@eprint@#1:#2::\@nil}
\def\mn@eprint@arXiv#1{\href {http://arxiv.org/abs/#1} {{\tt arXiv:#1}}}
\def\mn@eprint@dblp#1{\href {http://dblp.uni-trier.de/rec/bibtex/#1.xml}
  {dblp:#1}}
\def\mn@eprint@#1:#2:#3:#4\@nil{\def\@tempa {#1}\def\@tempb {#2}\def\@tempc
  {#3}\ifx \@tempc \@empty \let \@tempc \@tempb \let \@tempb \@tempa \fi \ifx
  \@tempb \@empty \def\@tempb {arXiv}\fi \@ifundefined
  {mn@eprint@\@tempb}{\@tempb:\@tempc}{\expandafter \expandafter \csname
  mn@eprint@\@tempb\endcsname \expandafter{\@tempc}}}

\bibitem[\protect\citeauthoryear{{Boyle} \& {Sclocco}}{{Boyle} \&
  {Sclocco}}{2019}]{Boyle2019}
{Boyle} J.,  {Sclocco} A.,  2019, in 2019 RFI Workshop - Coexisting with Radio
  Frequency Interference (RFI). pp~1--5,
  \mn@doi{10.23919/RFI48793.2019.9111722}

\bibitem[\protect\citeauthoryear{Buch, Bhatporia, Gupta, Nalawade, Chowdhury,
  Naik, Aggarwal  \& Ajithkumar}{Buch et~al.}{2016}]{Buch2016}
Buch K.~D.,  Bhatporia S.,  Gupta Y.,  Nalawade S.,  Chowdhury A.,  Naik K.,
  Aggarwal K.,   Ajithkumar B.,  2016, Journal of Astronomical Instrumentation,
  5, 1641018

\bibitem[\protect\citeauthoryear{Buch, Naik, Nalawade, Bhatporia, Gupta  \&
  Ajithkumar}{Buch et~al.}{2019}]{Buch2019}
Buch K.~D.,  Naik K.,  Nalawade S.,  Bhatporia S.,  Gupta Y.,   Ajithkumar B.,
  2019, Journal of Astronomical Instrumentation, 8

\bibitem[\protect\citeauthoryear{Cover \& Thomas}{Cover \&
  Thomas}{2006}]{Cover2006}
Cover T.~M.,  Thomas J.~A.,  2006, Elements of Information Theory.
John Wiley \& Sons, Inc., \mn@doi{10.1002/047174882X}

\bibitem[\protect\citeauthoryear{Dwyer}{Dwyer}{1983}]{Dwyer1983}
Dwyer R.,  1983, in ICASSP '83. IEEE International Conference on Acoustics,
  Speech, and Signal Processing. pp 607--610,
  \mn@doi{10.1109/ICASSP.1983.1172264}

\bibitem[\protect\citeauthoryear{{Ferrante}, {Masiero}  \& {Pavon}}{{Ferrante}
  et~al.}{2011}]{2011arXiv1103.5602F}
{Ferrante} A.,  {Masiero} C.,   {Pavon} M.,  2011, \mn@doi [arXiv e-prints]
  {10.48550/arXiv.1103.5602}, \href
  {https://ui.adsabs.harvard.edu/abs/2011arXiv1103.5602F} {p. arXiv:1103.5602}

\bibitem[\protect\citeauthoryear{Flanagan}{Flanagan}{1972}]{Flanagan1972}
Flanagan J.~L.,  1972, Speech Analysis, Synthesis and Perception..
Springer-Verlag, New York

\bibitem[\protect\citeauthoryear{Ford \& Buch}{Ford \& Buch}{2014}]{Ford2014}
Ford J.~M.,  Buch K.~D.,  2014, in 2014 {IEEE} Geoscience and Remote Sensing
  Symposium. {IEEE}, pp 231--234, \mn@doi{10.1109/IGARSS.2014.6946399}, \url
  {http://ieeexplore.ieee.org/document/6946399/}

\bibitem[\protect\citeauthoryear{{Foster}, {Wolszczan}  \& {Camilo}}{{Foster}
  et~al.}{1993}]{1993ApJ...410L..91F}
{Foster} R.~S.,  {Wolszczan} A.,   {Camilo} F.,  1993, \mn@doi [\apjl]
  {10.1086/186887}, \href
  {https://ui.adsabs.harvard.edu/abs/1993ApJ...410L..91F} {410, L91}

\bibitem[\protect\citeauthoryear{Gary, Liu  \& Nita}{Gary
  et~al.}{2010}]{gary2010}
Gary D.~E.,  Liu Z.,   Nita G.~M.,  2010, Publications of the Astronomical
  Society of the Pacific, 122, 560

\bibitem[\protect\citeauthoryear{Iglewicz \& Hoaglin}{Iglewicz \&
  Hoaglin}{1993}]{Iglewicz1993}
Iglewicz B.,  Hoaglin D.~C.,  1993, How to Detect and Handle Outliers.
ASQC Quality Press

\bibitem[\protect\citeauthoryear{Lorimer \& Kramer}{Lorimer \&
  Kramer}{2005}]{Lorimer12}
Lorimer D.,  Kramer M.,  2005, Handbook of pulsar astronomy.
Cambridge University Press, New York

\bibitem[\protect\citeauthoryear{{Lorimer}, {Bailes}, {McLaughlin}, {Narkevic}
  \& {Crawford}}{{Lorimer} et~al.}{2007}]{lbm+07}
{Lorimer} D.~R.,  {Bailes} M.,  {McLaughlin} M.~A.,  {Narkevic} D.~J.,
  {Crawford} F.,  2007, \mn@doi [Science] {10.1126/science.1147532}, \href
  {http://adsabs.harvard.edu/abs/2007Sci...318..777L} {318, 777}

\bibitem[\protect\citeauthoryear{McLaughlin et~al.,}{McLaughlin
  et~al.}{2006}]{mclaughlin2006transient}
McLaughlin M.~A.,  et~al., 2006, Nature, 439, 817

\bibitem[\protect\citeauthoryear{Morello, Barr, Stappers, Keane  \&
  Lyne}{Morello et~al.}{2020}]{Morello_2020}
Morello V.,  Barr E.~D.,  Stappers B.~W.,  Keane E.~F.,   Lyne A.~G.,  2020,
  Monthly Notices of the Royal Astronomical Society, 497, 4654

\bibitem[\protect\citeauthoryear{Moulin \& Veeravalli}{Moulin \&
  Veeravalli}{2019}]{Moulin2020}
Moulin P.,  Veeravalli V.~V.,  2019, Statistical Inference for Engineers and
  Data Scientists.
Cambridge University Press, New York, NY, USA

\bibitem[\protect\citeauthoryear{Nita \& Gary}{Nita \& Gary}{2010}]{nita2010}
Nita G.~M.,  Gary D.~E.,  2010, Monthly Notices of the Royal Astronomical
  Society: Letters, 406, L60

\bibitem[\protect\citeauthoryear{Nita, Gary, Liu, Hurford  \& White}{Nita
  et~al.}{2007}]{nita2007}
Nita G.~M.,  Gary D.~E.,  Liu Z.,  Hurford G.~J.,   White S.~M.,  2007,
  Publications of the Astronomical Society of the Pacific, 119, 805

\bibitem[\protect\citeauthoryear{Nita, Hickish, MacMahon  \& Gary}{Nita
  et~al.}{2016}]{nita2016}
Nita G.~M.,  Hickish J.,  MacMahon D.,   Gary D.~E.,  2016, \mn@doi [Journal of
  Astronomical Instrumentation] {10.1142/S2251171716410099}, 5, 1641009 (16
  pages)

\bibitem[\protect\citeauthoryear{Nita, Keimpema  \& Paragi}{Nita
  et~al.}{2019}]{nita2019}
Nita G.~M.,  Keimpema A.,   Paragi Z.,  2019, Journal of Astronomical
  Instrumentation, 8

\bibitem[\protect\citeauthoryear{{Ramey}, {Joslyn}, {Prestage}, {Lam},
  {Hawkins}, {Blattner}  \& {Whitehead}}{{Ramey}
  et~al.}{2019}]{Real-timeRFI:eramey}
{Ramey} E.,  {Joslyn} N.,  {Prestage} R.,  {Lam} M.,  {Hawkins} L.,  {Blattner}
  T.,   {Whitehead} M.,  Washington University in St. Louis, 2019, Seniors
  Honor paper/Undergraduate Thesis

\bibitem[\protect\citeauthoryear{Saroff}{Saroff}{2023}]{saroff}
Saroff D.,  2023, PhD thesis, Rochester Institute of Technology

\bibitem[\protect\citeauthoryear{Shannon}{Shannon}{1948}]{Shannon1948}
Shannon C.~E.,  1948, Bell System Technical Journal, 27, 379

\bibitem[\protect\citeauthoryear{Shapiro \& Wilks}{Shapiro \&
  Wilks}{1965}]{Shapiro1965}
Shapiro S.~S.,  Wilks M.~B.,  1965, Biometrika, 52, 591

\bibitem[\protect\citeauthoryear{Shen, Hung  \& Lee}{Shen
  et~al.}{1998}]{Shen1998RobustEE}
Shen J.-L.,  Hung J.-W.,   Lee L.,  1998, in ICSLP.

\bibitem[\protect\citeauthoryear{Staelin}{Staelin}{1969}]{1969Fast_Folding_Algorithm}
Staelin D.~H.,  1969, IEEE Proceedings, 57, 724

\bibitem[\protect\citeauthoryear{Taylor, Denman, Bandura, Berger, Masui,
  Renard, Tretyakov  \& Vanderlinde}{Taylor et~al.}{2019}]{Taylor2019}
Taylor J.,  Denman N.,  Bandura K.,  Berger P.,  Masui K.,  Renard A.,
  Tretyakov I.,   Vanderlinde K.,  2019, \mn@doi [Journal of Astronomical
  Instrumentation] {10.1142/S225117171940004X}, 8, 1940004 (9 pages)

\bibitem[\protect\citeauthoryear{{Thornton} et~al.,}{{Thornton}
  et~al.}{2013}]{2013Sci...341...53T}
{Thornton} D.,  et~al., 2013, \mn@doi [Science] {10.1126/science.1236789},
  \href {https://ui.adsabs.harvard.edu/abs/2013Sci...341...53T} {341, 53}

\makeatother
\end{thebibliography}
\bsp	
\label{lastpage}
\end{document}